# Revealing Nanoscale Confinement Effects on Hyperbolic Phonon Polaritons with an Electron Beam

*Andrea Konečná[1*], Jiahan Li[2], James H. Edgar[2], F. Javier García de Abajo[1,3], Jordan A. Hachtel[4*]*
[1] ICFO-Institut de Ciencies Fotoniques, The Barcelona Institute of Science and Technology, 08860 Castelldefels (Barcelona), Spain
[2] Tim Taylor Department of Chemical Engineering, Kansas State University, Manhattan, Kansas 66506, USA
[3] ICREA-Institució Catalana de Recerca i Estudis Avançats, Passeig Lluís Campanys 23, 08010, Barcelona, Spain
[4] Center for Nanophase Materials Sciences, Oak Ridge National Laboratory, Oak Ridge, TN 37831 USA
[*] Correspondence to: andrea.konecna@icfo.eu
[*] Correspondence to: hachtelja@ornl.gov

**Abstract**

Hyperbolic phonon polaritons (HPhPs) in hexagonal boron nitride (hBN) enable the direct manipulation of mid-infrared light at nanometer scales, many orders of magnitude below the free-space light wavelength. High resolution monochromated electron energy-loss spectroscopy (EELS) facilitates measurement of excitations with energies extending into the mid-infrared while maintaining nanoscale spatial resolution, making it ideal for detecting HPhPs. The electron beam is a precise source and probe of HPhPs, that allows us to perform novel experiments to observe nanoscale confinement in HPhP structures and directly extract hBN polariton dispersions for both modes in the bulk of the flake and modes along the edge. Our measurements reveal technologically important non-trivial phenomena, such as localized polaritons induced by environmental heterogeneity, enhanced and suppressed excitation due to two-dimensional interference, and strong modification of high-momenta excitations of edge-confined polaritons by nanoscale heterogeneity on edge boundaries. Our work opens exciting prospects for the design of real-world optical mid-infrared devices based on hyperbolic polaritons.

**Main Text**

The confinement and manipulation of light through localized plasmon- and phonon-polaritons is widely used to control light in the visible and near-infrared spectral regimes at the nanoscale[1–3]. The less developed mid-infrared (mid-IR) range holds great potential for new applications[4–6], but is still limited by the lack of suitable materials systems in which low-energy excitations can be easily excited and manipulated. Hexagonal boron nitride (hBN) is one such material, which due to its dielectric anisotropy, possesses two Reststrahlen bands that support hyperbolic phonon polaritons (HPhPs) that confine mid-IR light to nanometer-scales and allow it to propagate over micron-scales[7,8].

Only recently have optical analysis techniques achieved the spatial-resolution necessary to access HPhPs directly at the nanoscale. Scanning near-field optical microscopy (SNOM) is extensively used to probe the localization and dispersions of HPhPs at nano-length-scales[9], as studying other types of mid-IR light/matter interactions[10,11] in polaritonic nanostructures, heterostructures, and metamaterials[12–15]. It's limiting factor of SNOM, however, is the spatial resolution, which is typically limited to a few tens of nanometers by the size of the probe tip. Recently, electron energy-loss spectroscopy (EELS) in the scanning transmission electron microscope (STEM) has emerged as an alternative, more precise technique, as it can access the mid-IR with sub-Ångstrom spatial resolution and as low as 3 meV spectral resolution[16–18]. The electron beam can access a number of low-energy excitations including localized phonon modes[19–22], molecular vibrations[23–26], and the HPhPs[27–29]. Moreover, the precise control of the probe enables on-demand excitation of polarization-dependent modes[30], directly induced localized quasiparticle coupling[31–33], and measurements of polariton dispersions[29,34–36]. Consequently, while EELS lacks the spectral resolution of SNOM, its combined spatial- and spectral resolution provides access into phenomena that cannot be explored in optical experiments.

Here, we discover several unique aspects of the nanoscale polaritonic response in hBN nanostructures by spectrally imaging HPhPs using monochromated STEM-EELS. A key property of HPhPs is geometrical confinement, and therefore, we examine the distinction between (1) HPhPs confined to the bulk of an exfoliated hBN flake and (2) HPhPs confined to the edge. Edge polaritons have been identified using EELS[29] and measurement of their dispersion have been performed in SNOM[37], but here we provide an EELS dispersion measurement of the edge polariton. The use of EELS is critical, as the high-spatial-resolution of the STEM accesses the high-wave-vector range of the edge polariton dispersion, revealing an extreme sensitivity to morphology of the flake wall. In addition, we find localization of polariton modes induced by either material contacting the surface of the hBN flake and the flake geometry itself. For all of our measurements, we carefully model the behavior of the polaritons using idealized systems and demonstrate that nanoscale heterogeneity in the sample is responsible for the formation of the localized polariton modes.

A key aspect of the EELS analysis is that the beam is both the source and the probe, enabling control of excitations through beam positioning. **Figure 1** illustrates how the probe position on and around an hBN flake changes the excitation of HPhPs and the EEL spectra. Panels in Fig. 1a-i show the electromagnetic simulations of the polariton response at 180 meV, 185 meV, and 190 meV for probe positions 100 nm outside the flake, at the flake edge, and 100 nm inside the flake. For all probe positions and energies the beam excites polaritons that propagate into the flake in all directions, as well as along the edge, even when the beam does not directly intersect the flake due to the 'aloof' effect[38]. The primary difference between probe positions is the relative intensity of the bulk vs. edge excitation, and the primary difference between energies are the wavelengths of the excitations in the flake. It is also important to note that at all energies the wavelength of the edge excitation is visibly shorter than the bulk excitation.

These effects manifest in the EELS line profile of an hBN flake edge. A high angle annular dark field (HAADF) image of a flake edge is shown in Fig. 1j, with the EEL line profile in Fig. 1k. The EELS has a peak at ~176 meV that is excited everywhere, corresponding to the bulk polariton, but when the probe is close to the edge, higher energy losses up to ~185 meV are observed, corresponding to the edge polariton. Only when the probe is within ~10 nm of the edge wall is there a sufficiently strong excitation of the edge polariton to resolve it over the bulk polariton signal, but as illustrated in the simulations it is still being excited at all probe positions. The high degree of localization of the edge-polariton is surprising given that the aloof excitations are normally delocalized on the scale of hundreds of nm for comparable energy levels[39]. However, the result follows naturally from the high degree of confinement of the edge-polariton to the flake wall, and the geometrical constraints on momentum transfer from the beam. When the beam is at the edge of the sample, it can transfer a large amount of momentum along the flake-edge, enabling efficient excitation of the high-momentum (and hence high-energy-loss) branch of the polariton dispersions. Then, as the beam moves further away, only lower-momentum (lower-energy-loss) excitations are detected. The effect is most visible for the edge polariton, is also present in the aloof excitation of the bulk polariton, which continually red-shifts as the probe moves further away from the sample (see Supplementary Figure 1).

We also note that peaks at 160 meV and 200 meV are observed when the beam does intersect the flake, corresponding to the TO and LO phonons, respectively. These excitations are only accessed through impact scattering, and are not part of the polaritonic response, but they are still important as they define the limits of the Reststrahlen band in which the HPhPs are active.

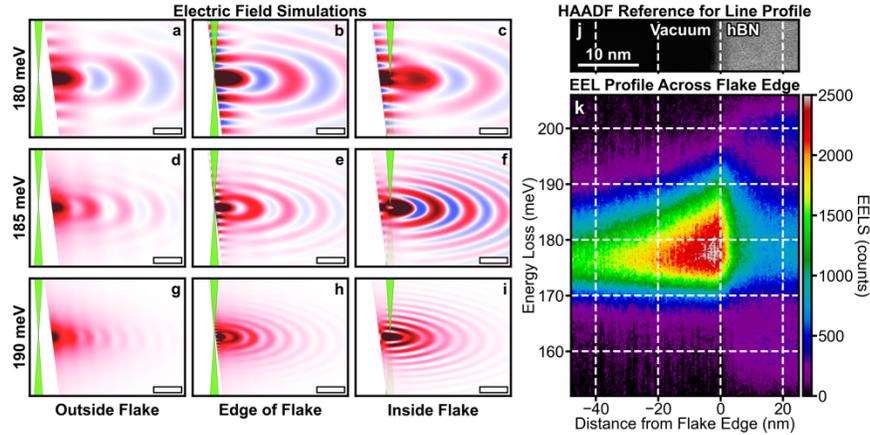

**Figure 1. EELS Excitation of Hyperbolic Phonon Polaritons (HPhPs) at the Edge of an hBN Flake. (a-i)** Electric field simulations of the polaritonic response in an hBN flake for different probe positions: 100 nm outside the flake (a,d,g), at the flake edge (b,e,h), and 100 nm inside the flake (c,f,i) at energy losses of 180 meV (a-c), 185 meV (d-f), and 190 meV (g-i). Scale bars = 200 nm. **(j)** HAADF reference image for line profile across hBN flake edge. **(k)** Measured EELS line profile across a flake edge demonstrating highly localized excitation of the edge-confined polaritons.

The momentum selectivity afforded through careful probe positioning can go a step further to directly acquire HPhP dispersions with polariton interferometry[27,29,34–36], also frequently used in SNOM[9,11,12]. This is exemplified for the bulk polariton dispersion in **Figure 2**. The experiment is shown schematically in Fig. 2a: a line profile is acquired normal to the flake edge, and polaritons that propagate directly towards the edge of the flake can reflect back resulting in interference. Such a line profile is plotted in Fig. 2b, and as the probe moves from deep (~200 nm) into the flake towards the edge, a dispersive peak corresponding to constructive interference with reflected polaritons emerges. As a result, each probe position approximately corresponds to a specific wavelength for which interference is maximized. Thus, by fitting the peak energy as a function of distance from the edge, the entire dispersion is measured (details of the fitting process expanded in Supplementary Discussion 1).

The interference peaks also emerge in our numerical modeling. Figures 2c and 2d show the simulated EELS line profile, both broadened to match our experimental EELS energy resolution of 6 meV (c) and as directly obtained from the simulation (d). The broadened profile qualitatively and quantitatively matches with the EELS experiment, demonstrating that EELS accurately measures the bulk polariton interference. The full simulation (Fig. 2d) also reveals that the broad tail extending out to the upper limit of the Reststrahlen band (the LO phonon) at 200.1 meV in the experimental and broadened-simulation profiles actually consists of the higher-order interference maxima that cannot be individually identified at 6 meV energy resolution.

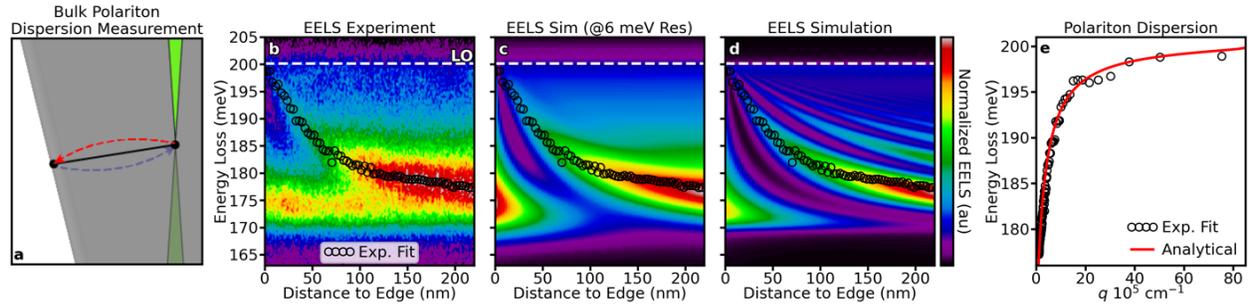

**Figure 2. Hyperbolic Polariton Dispersion Measurement in hBN Flakes. (a)** Schematic of polariton interferometry for a bulk HPhP dispersion measurement. Polaritons reflected from the edge create constructive interference at different wave vectors of the HPhP dispersion. **(b)** Experimental EELS line profile and fit of first-order interference peak (black circles). **(c,d)** Simulated EELS line profiles broadened to match the 6 meV experimental energy resolution (c) and at the native resolution of the calculation (d). **(e)** The dispersion is achieved by transforming the distance to the edge in the EELS measurement considering $\phi_R = \frac{\pi}{2}$. The experimental dispersion compared to the analytical expression, showing a match out to high $q$ values.

We now transform the distances on the *x*-axis of the interference peak to dispersion values, using the transformation $q = \frac{2\pi - \phi_R}{2x}$, where $q$ is the polariton wave vector, $x$ is the distance to the edge and $\phi_R$ is a reflection phase that modifies the polariton upon reflection. Here, we use $\phi_R = \frac{\pi}{2}$, which presents the best match between theory and experiment (Supplementary Figure 2 and Discussion 2). The transformed experimental dispersion is directly compared the analytical expression, Supplementary Equation 1, in Fig. 2e, achieving excellent agreement almost all the way out to $10^7$ cm$^{-1}$.

We can also apply the same interferometry technique to measure the dispersion of edge-confined polaritons, as shown in **Figure 3**. Here, an aloof line profile is acquired moving along the edge of the flake towards a sharp corner, enabling excited edge polaritons to propagate to the corner, reflect, and interfere with one another (see schematic in Fig. 3a). The experimental and simulated (with and without broadening) EELS line profiles for the edge polariton are shown in Fig. 3b-d, respectively, with the experimental fitted peak values plotted as black crosses in all three panels. While a dispersive peak is clearly measured in both experiment and simulation, the agreement with theory here is far worse than it is for the bulk polariton in Fig. 2.

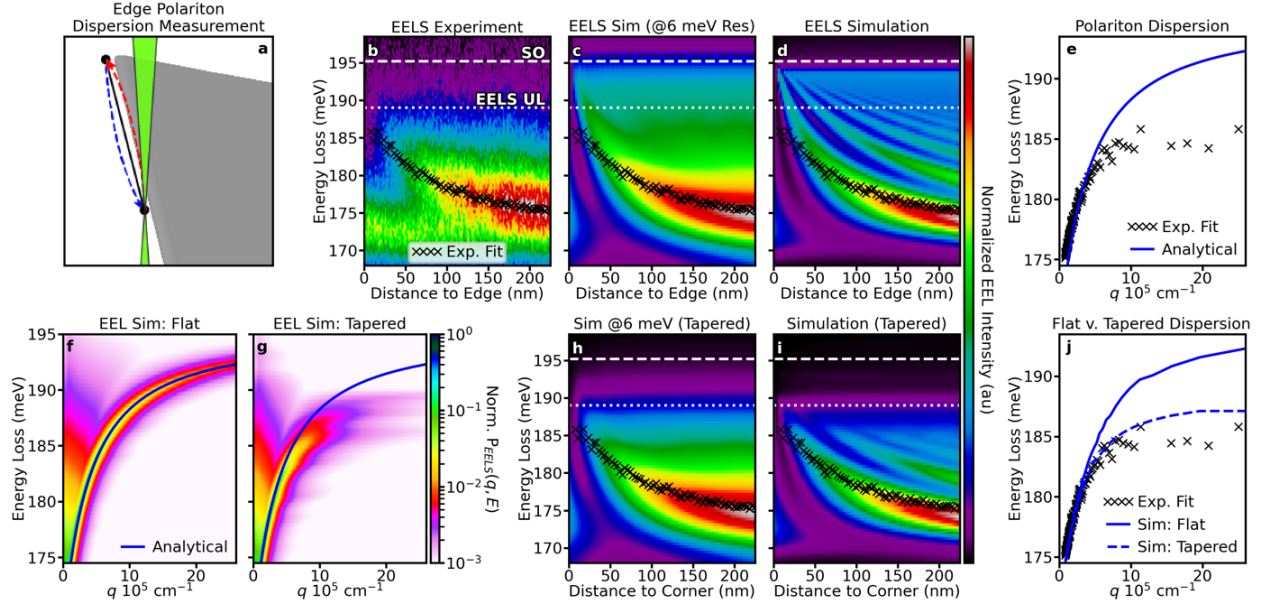

**Figure 3. Polariton Dispersion Measurement at Heterogeneous Edges in hBN Flakes. (a)** Schematic of polariton interferometry for an edge HPhP dispersion measurement. Polaritons reflected from the corner along the edge produced by HPhP interference. **(b-d)** Experimental and simulated EELS line profiles: experiment (b), simulated and broadened to match EELS energy resolution (c), and the actual simulation (d). In (b-d) the upper limit of the measured EELS (at 189 meV) is marked with a dotted line and the theoretical upper limit (the SO phonon at 195.2 meV) is marked with a dashed line, demonstrating that EELS and simulated line scans have significantly different upper limits. **(e)** EELS dispersion transformed with $\phi_R = \frac{3\pi}{4}$ compared to the analytical expression for the edge polariton, exhibiting significant discrepancies. **(f,g)** Momentum-resolved EELS simulations for a 90º flat (f) and 45º tapered (g) edge, illustrating that high-$q$ excitations are suppressed in the tapered edge (more closely resembling the excitations measured in the experimental line profile). **(h,i)** Broadened (h) and actual (i) line profiles simulated for a tapered edge flake, producing a much better match to the experimental upper limit. **(j)** Comparison of the experimental and simulated (for both flat and tapered edges) edge HPhP dispersions.

Perhaps most importantly, the upper limit of the experiment appears to be ~189 meV, while it is at 195.2 meV in the simulations, corresponding to the energy of the high-$q$ SO phonon[27]. Additionally, the EELS fits deviate from the simulated first-order peaks as the probe gets closer to the corner. The experimental dispersion is plotted in Fig. 3e (using $\phi_R = \frac{3\pi}{4}$ for the transformation, see Supplementary Figure 3 and Discussion 2 for more details), and when compared to the analytical expression, Supplementary Equation S4, a significant deviation is observed. We attribute the effect to the nanoscale morphology of the flake wall. The above calculations treat the edge as a perfectly flat face, while real hBN flakes typically have significant heterogeneity and thickness variation along the edge. The bulk HPhPs in the interior of the flake exist in a nominally uniform environment, but the edge HPhPs are confined directly to this heterogeneous region. Edge heterogeneity can be partly included in the simulations by changing the model system from a flat 90º wall to a tapered 45º wall to account for the step-like thickness gradient normally found in exfoliated flakes. Simulations of the $q$-resolved EELS probability of an excitation at a flat and tapered hBN flake edge are shown in Figs. 3f and 3g, respectively. The flat edge probability exhibits high intensity directly along the analytical dispersion, but the tapered edge probability, is both suppressed and red-shifted from the analytical dispersion, especially at high-$q$, similar to the experiment.

The simulated (with and without broadening) EELS line profiles for the tapered edge are shown in Figs. 3h and 3i. Here, the simulated upper limit matches to the experimental upper limit, and the fitted experimental peaks match the simulated first-order peak much more closely. The dispersions at the flat and tapered edge are determined by fitting the dispersive peaks in Figs. 3c and 3h, respectively, and are compared to the experimental dispersion in Fig. 3j. The flat-edge simulation resembles the analytical dispersion in Fig. 3e, while the tapered-edge simulation resembles the experimentally-retrieved dispersion. The comparison emphasizes the power of EELS as a complementary technique to SNOM, as SNOM typically can only probe $q$ values up to a few times $10^5$ cm$^{-1}$.[9,29,37] However, as revealed by Fig. 3e, the deviations between the analytical and experimental edge polariton expressions only become significant above $10^6$ cm$^{-1}$, necessitating the high-$q$ interferometry in EELS with a well-defined focused probe.

The bulk polaritons can also be modified by subtle nanoscale environmental and geometric heterogeneity. In **Figure 4**, we consider the effect of materials contacting the surface of the hBN flakes. All our samples are supported on lacey carbon TEM grids, nominally thought to only support the flake. However, polaritonic modes are so sensitive to the surrounding environment[34] that the lacey carbon directly induces localized polariton modes in the flake. A schematic of our experiment is shown in Fig. 4a. We find a hole in the lacey carbon underneath a large flake (far from the edges) and probe the hole as a function of distance from the carbon edge. Figure 4b shows the HAADF image of the region used for the hyperspectral acquisition, centered around a ~400 nm diameter hole in the lacey carbon. For reference, the inset contains a low-magnification STEM image of the flake with the selected hole highlighted.

The hyperspectral dataset is partitioned into regions at different average distances $D$ away from the carbon (i.e., the white ring in Fig. 4b). We use 10 different distances (highlighted by different colors in Fig. 4c) and plot the average EELS signal from each in Fig. 4d. With no carbon/hBN interaction the EELS would be uniform in all 10 regions, which is clearly not the case. The intensity of the bulk polariton peak increases by ~20% compared to the edge and it blueshifts slightly (173.3 meV to 174.6 meV). In addition, at the outmost region ($D$=13 nm) there is a shoulder at about 190 meV, which redshifts to 185 meV for the $D$=51 nm region, and then further up the edge of the dominant peak as the probe gets closer to the center of the hole. The effect can be understood by considering different pathways for polaritons excited by the beam: they can propagate past the carbon, or they can be reflected off it like off the edge of a disk. Thus, we describe the EELS as a combination of the two effects: infinite slab excitations (Fig. 4e) and finite disk excitations (Fig. 4f).

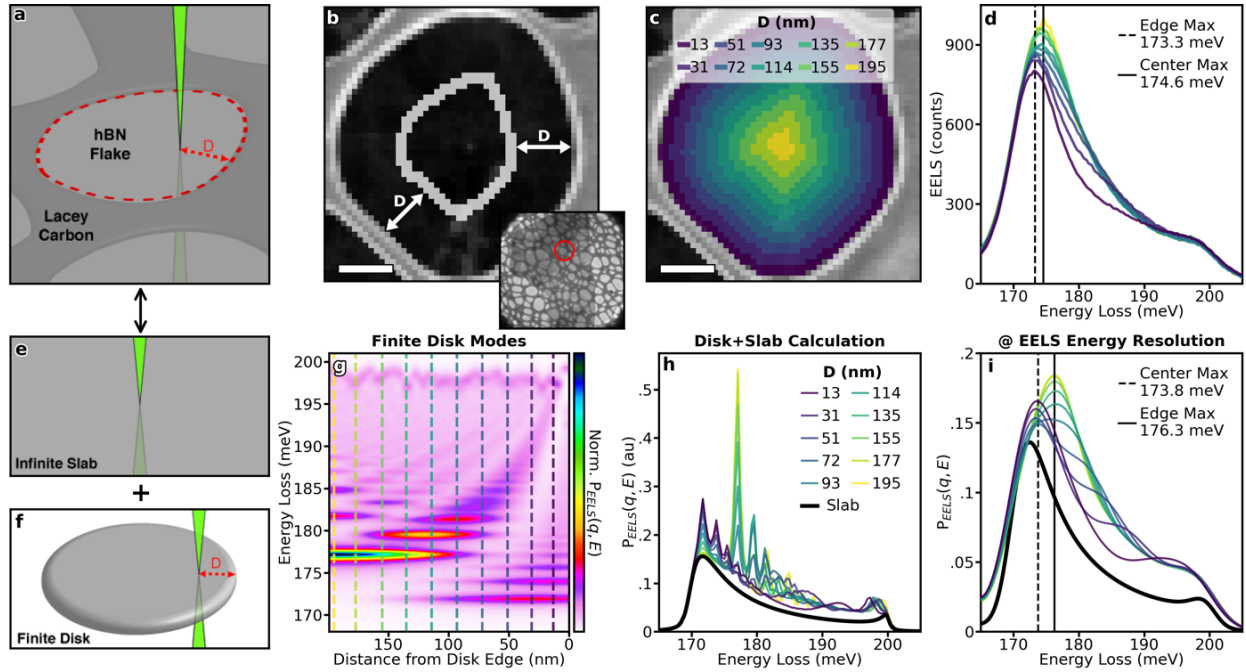

**Figure 4. Localized Polariton Modes Induced by Environmental Heterogeneity. (a)** Schematic of experiment to probe polariton interaction with supporting lacey carbon. **(b)** HAADF reference image of the region of interest Inset: low-magnification STEM image of the entire hBN flake with the hole marked in red. **(c)** Different regions of the selected hole in lacey carbon classified by the average distance $D$ from carbon edge. **(d)** Spectra acquired from regions highlighted in (c). Scale bars=100 nm. **(e,f)** Spectra in (d) can be described as a combination of the infinite slab response expected in the center of a large flake (e) with finite-disk-like modes induced by the lacey carbon (f). **(g)** Calculated modes for a finite hBN disk with 200 nm radius. $D$ values from (c) and (d) are plotted as colored dashed lines. **(h)** Simulated EELS spectra for the same $D$ values as in (c) and (d), obtained by combining the spectrum of an infinite slab signal (black), with the $D$-dependent disk modes (color). **(i)** Spectra in (h) broadened to match the EELS energy resolution, showing qualitative agreement with experiment in (d).

The calculated localized polariton modes of a 400 nm diameter finite hBN disk are plotted in Fig. 4g. We detect multipole modes excited at the edge of the disk, breathing modes excited at the center, and *hybrid* modes emerging as combination of those two (analogous to localized surface plasmon modes[40,41]). We select finite disk spectra to match the average $D$ from the regions in Fig. 4c, then add them to the infinite slab spectrum, and finally plot the combination in Fig. 4h. The results after broadening with the 6 meV resolution of our microscope are plotted in Fig. 4i.

The superposition of slab and disk excitations reproduces the major effects in the experiment, exhibiting enhanced intensity and a small blueshift of the peak energy at the center of the flake, which can now be attributed to the different localizations of the multipole and breathing modes. Moreover, Fig. 4g shows that the *hybrid* modes have a dispersive change in energy as the probe gets closer to the edge, explaining the secondary position-dependent peak in the EELS data, which is also reproduced in the calculations. The localized disk modes can be further visualized using linear unmixing techniques as shown in the Supplementary Figure 4.

The geometry of the flake boundaries also has a dominant effect on the bulk polariton response. Here, we return to the flake used in Figs. 2 and 3, which possessed a sharp corner, and acquire a hyperspectral image from the entire region (**Figure 5**). The probe interacts with both edges of the flake simultaneously, as shown schematically in Fig. 5a. In Fig. 5b-e, we show experimental energy-filtered EELS images for 0.5 meV wide windows centered at 180, 185, 190, and 195 meV, respectively, capturing the 2D interference pattern at the hBN corner. The corresponding EELS simulations are shown in Fig. 5f-i.

The latter show a much larger degree of interference detail (especially at higher energies) compared to the experimental EELS with limited energy resolution. However, several key aspects of the interference patterns are still clearly visible in the EELS experiment. Notably, interference minima forming dark fringes along the flake edges are seen in experiment and are similar in width and energy dependence to the dark fringes in the simulations. The simulations also show higher order interference minima and maxima deeper into the flake, increasing in number at the higher energies, which is mostly blurred out at the 6 meV energy resolution of EELS. However, even these higher order fringes are present in some circumstance circumstances, note that in both the 185 meV and 190 meV experimental images bright spots/fringes appear from higher-order positive interference. Additionally, the triangle-like geometry generates a localized-edge-polariton corner mode that can are the most pronounced in the 180 meV and 185 meV simulations as a bright spot on the edge of the flake (further details are provided in Supplementary Figure 5 and Discussion 3), while faint, comparable bright spots are present at the same position in the EELS image at 180 meV (detectable features highlighted directly in Supplementary Figure 6).

Lastly, it is important to note far away from the corner the signal is dominated by the interaction with the nearest edge, however, close to the tip (within 100 nm) the polaritonic response is strongly influenced by both edges simultaneously and the full 2D geometry is needed to understand the response. This 2D interference is observed in the experiment almost as clearly as in the theory, demonstrating the ability to use STEM-EELS to map constructive and destructive interference in nanostructured polaritonic materials.

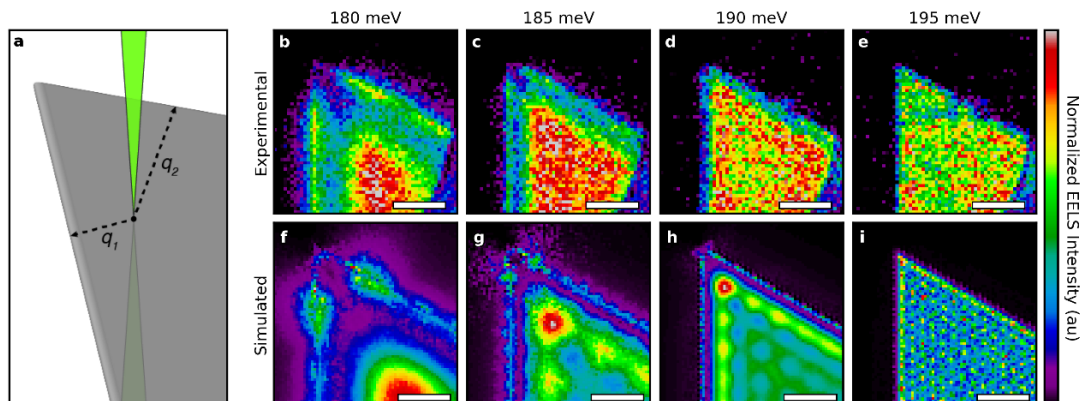

**Figure 5. Phonon polariton mapping in the Corner of a hBN Flake. (a)** Schematic of the experiment geometry. **(b-e)** Experimental measurements. **(f-i)** Simulations. The maps show the simultaneous existence of both bulk and edge HPhPs, producing both positive and negative interference along the surface and at the edges, respectively, and a localized mode emerging close to the corner tip. Scale bars=100 nm.

Conclusion

We have revealed important characteristics of hyperbolic phonon polaritons in hBN flakes by controllably exciting and detecting these modes using a nanoscale electron probe. The polaritons are extremely sensitive to the surrounding environment and geometry, which enable us to probe a wide range of polariton properties including the dispersions of edge and bulk HPhPs, localized modes induced by the surrounding environment and the flake geometry, as well 2D interference patterns in nanostructured flakes. Our interferometry measurements show that, at high wave vector, nanoscale heterogeneity at the flake edge significantly modifies the real dispersion of the edge-confined polaritons, in a way that cannot be resolved by traditional infrared optical experiments. The bulk-confined polaritons are also highly sensitive to local nanoscale heterogeneity and can create 2D interference that suppresses or enhances the excitation of HPhPs at nano-length scales. Interaction of such heterogeneities with material gives rise to localized modes determined by morphology. The precise control and measurement afforded by monochromated EELS opens the possibility to design complex devices on nanostructured hBN flakes and both measure and trigger their responses with precise control of the electron beam.

Methods

*Sample Growth*: hBN crystals were grown at atmospheric pressure from a molten iron-chromium solution using a $^{10}$B isotopically enriched boron source[42]. Flakes were then chemically exfoliated and transferred onto lacey carbon TEM grids, where the flakes were only supported by the thin carbon web. The sizes of the flakes varied from tens of microns to tens of nanometers in diameter and a few layers to a few hundred nanometers in thickness. The flakes used in experiment were chosen as the ones with the most uniform thickness and cleanest edges, as determined from low magnification STEM imaging for monochromated EELS analysis.

*STEM-EELS Experiment*: All EELS spectra were acquired on a Nion aberration-corrected high energy resolution monochromated EELS-STEM (HERMES$^{TM}$) equipped with the Nion Iris Spectrometer[43], operated at 30 kV with a convergence angle of 27 mrad and a collection angle of 25 mrad. All spectra were measured to have an energy resolution, as measured by the ZLP full-width at half-maximum between 5.5 meV and 6.5 meV. In all acquisitions, a power law background was used to fit the spectral regions directly before (120 meV – 140 meV) and after (220 meV –250 meV) the Rehstrahlen band and then subtracted to remove contributions from the ZLP tail.

Acquisition details are as follows: The line profile in Fig. 1 consisted of 150 spectra, with a calibration of 0.5 nm per pixel and a dispersion of 0.22 meV per spectrometer channel and an acquisition time of 3 seconds per spectrum. The line profiles in Fig. 2 and Fig. 3 are both 2D spectrum images that are binned in the axis normal to dispersive direction to create line profiles. For the bulk polariton measurement a 9x192 pixel spectrum image was acquired that was subsequently vertically binned across all nine pixels, with a calibration of 1.5625 nm per pixel, a dispersion of 0.40 meV per spectrometer channel, and an acquisition time of 128 ms per pixel. For the edge polariton measurement, a 163x20 pixel spectrum image was acquired and binned across only the two pixels closest to the edge extending into the flake (region for binning shown in Supplementary Figure 4). It was found that including pixels in the edge of the flake for horizontal binning did not vary the measured dispersion significantly, and only had the effect of including small contributions from the LO phonon, which made the upper limit of the measured dispersion more difficult to quantify. The spectrum image had a calibration of 1.5703 nm per pixel, a dispersion of 0.40 meV per spectrometer channel, and an acquisition time of 500 ms per pixel. For Fig. 3, a 50x50 pixels spectrum image was acquired with a calibration of 10 nm per pixel, a dispersion of 0.40 meV per spectrometer channel, and an acquisition time of 400 ms per pixel. For Fig. 4, a 64x64 pixel spectrum image was acquired, with a calibration of 4 nm per pixel, a dispersion of 0.40 meV per spectrometer channel, and an acquisition time of 400 ms per pixels.

We have also included videos of the spectrum images for the bulk polariton, edge polariton, finite disk, and corner analyses in supplemental, where a normalized 2D intensity is shown at frame-by-frame for all the energies in the Reststrahlen band in the Supplementary information.

*Calibration and Validation of Datasets*: Many factors can strongly influence the measured polariton dispersion, which need to be properly accounted for. In this analysis, we have included the specific pixel in the dataset deemed as the edge of the flake (a non-trivial assignment), the calibration of the spectrometer, which is difficult to carry out more accurately than around 1% at the high dispersions used in monochromated EELS, and the reflection phase. We calibrate our measurements to the LO phonon at 200.1 meV in the BN flakes, the energy of which is not dependent on flake thickness or geometry. For the dataset in Fig. 1, we found that a dispersion correction of 0.5% is needed. For the datasets in Fig. 2-5, which were all performed on the same day at the same dispersion, we found a needed dispersion correction of 1%. For the reflection phases and the edge pixel determination, which only applied to Figs. 2 and 3, we compared to the simulated line profiles to determine the correct values. This calibration and selection process is shown in detail in the Supplementary Discussion 2.

*Simulations*: All calculations were performed within the framework of classical macroscopic electrodynamics, which is known to reproduce very well the interaction of focused electron probes with optical excitations[44] in nanostructures. We used analytical expressions to calculate EELS spectra when dealing with an electron beam interacting with an infinite hBN slab. For more complex geometries (linear edge, corners, and disks), we used numerical solvers of the Poisson and Maxwell equations implemented in the commercial software Comsol Multiphysics. Further details on the formalism, analytical expressions, and models, as well as setting parameters for the numerical solvers, are described in the Supplementary Discussion 3.

Author Contributions

Monochromated EELS experiments and analysis were performed by JAH. Theoretical calculations and simulations were performed by AK and JGA. hBN flakes synthesized by JL and JHE. Experimental strategy was developed by AK and JAH. Manuscript and Supplementary Information were prepared by AK and JAH. All authors participated in the editing and revision of the manuscript.

Acknowledgements

Microscopy experiments conducted at the Center for Nanophase Materials Sciences, which is a DOE Office of Science User Facility using instrumentation within ORNL's Materials Characterization Core provided by UT-Battelle, LLC, under Contract No. DE-AC05- 00OR22725 with the DOE and sponsored by the Laboratory Directed Research and Development Program of Oak Ridge National Laboratory, managed by UT-Battelle, LLC, for the U.S. Department of Energy (JAH). AK and JGA acknowledges support from the European Research Council (Advanced Grant No. 789104eNANO), the European Commission (Horizon 2020 Grants No. FET-Proactive 101017720-EBEAM and No. FET-Open 964591-SMART-electron), and the Spanish MINECO (MAT2017-88492-R and SEV2015-0522). Support for hBN crystal growth from the Office of Naval Research (Award No. N00014-20-1-2474) is appreciated (JL and JHE).

Competing Interests

The authors declare no competing interests.

# Supplementary Information for
# Revealing Nanoscale Confinement Effects on Hyperbolic Phonon Polaritons with an Electron Beam

*Andrea Konečná[1*], Jiahan Li[2], James H. Edgar[2], F. Javier García de Abajo[1], Jordan A. Hachtel[3*]*

[1] ICFO-Institut de Ciencies Fotoniques, The Barcelona Institute of Science and Technology, 08860 Castelldefels (Barcelona), Spain
[2] Tim Taylor Department of Chemical Engineering, Kansas State University, Manhattan, Kansas 66506, USA
[3] ICREA-Institució Catalana de Recerca i Estudis Avançats, Passeig Lluís Campanys 23, 08010, Barcelona, Spain
[4] Center for Nanophase Materials Sciences, Oak Ridge National Laboratory, Oak Ridge, TN 37831 USA
[*]Correspondence to: andrea.konecna@icfo.eu
[*]Correspondence to: hachtelja@ornl.gov

**Supplementary Figures**

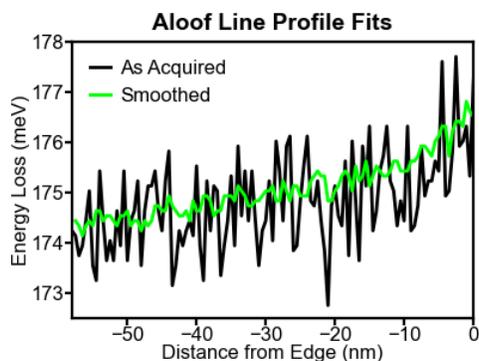

**Supplementary Figure 1. Redshifting Peak HPhP Energy.** As referenced in the main text, and observable here, when the probe is outside the flake (distance from edge < 0), the HPhP peak is still observed through aloof excitation. High-$q$ transfer is only possible in extremely close proximity to the electron beam, so as we move further away the high-$q$ (high energy) branch of the polariton dispersion is suppressed faster than the low-$q$ (low energy) branch, causing the peak to effectively redshift. The HPhP peak redshifts by over 2 meV in 50 nm due to the lack of excitation of high-$q$ polaritons. Further details of fitting process in Supplementary Discussion 1.

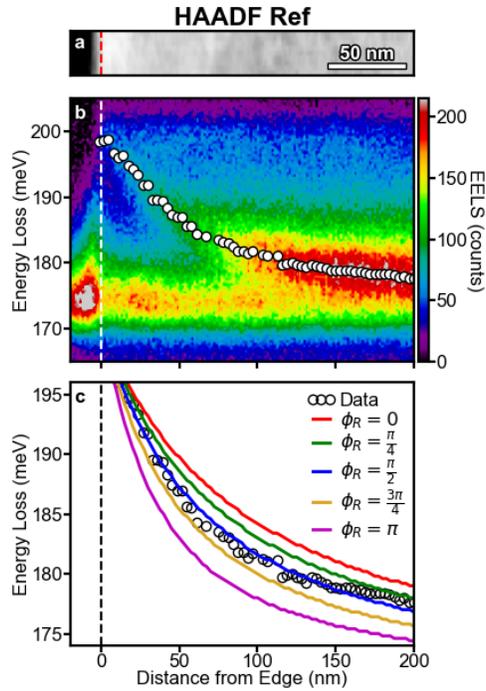

**Supplementary Figure 2. Comparison of Measured EELS Bulk Line Profile to Simulations with Differing Reflection Phases.** Using the process for fitting outlined in Supplementary Discussion 1, we fit the experimental line profile by comparison with simulated line profiles performed with reflection phases of $\phi_r = 0, \frac{\pi}{4}, \frac{\pi}{2}, \frac{3\pi}{4}, and\ \pi$. Panel (a) shows the HAADF reference of the edge of the flake used for the bulk polariton dispersion measurement. Panel (b) the EELS line profile with the fitted peaks originating from the interference. Panel (c) shows the fitted EELS peaks compared to the fits for the different simulations. For the bulk dispersion, $\phi_r = \frac{\pi}{2}$ provides the best match to the data. For additional details on the comparison to different reflection phase, see Supplementary Fig. 15.

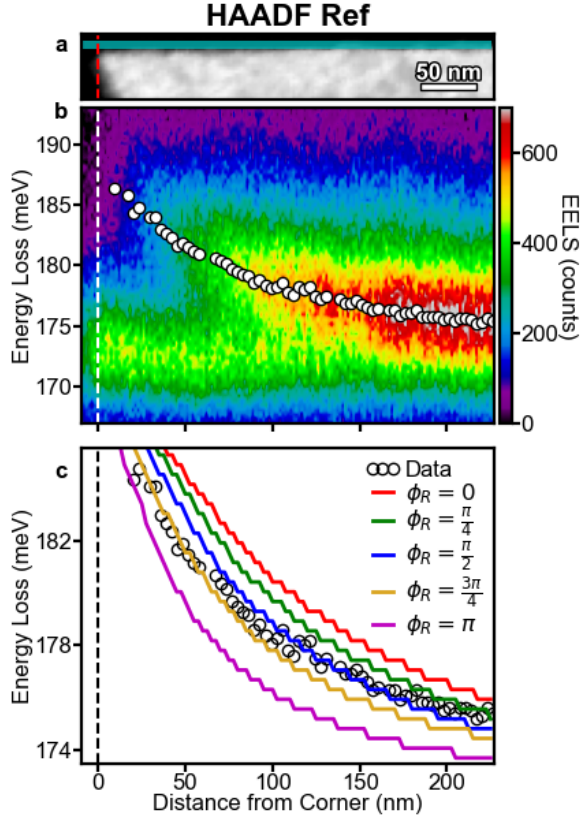

**Supplementary Figure 3. Comparison of Measured EELS Edge Line Profile to Simulations with Differing Reflection Phases.** The same comparison for the edge experimental line profile and the tapered edge simulated line profiles with $\phi_r = 0, \frac{\pi}{4}, \frac{\pi}{2}, \frac{3\pi}{4}$, and $\pi$. (a) HAADF reference image showing the corner more clearly, and the cyan regions demarcate the pixels used for the line profile. (b) EELS line profile with the EELS peak fits. (c) Fitted peaks compared to those obtained for the different simulations. Unlike in **Supplementary Fig. 2** for the bulk dispersion, there is not one value that exhibits an ideal match. We choose $\phi_r = \frac{3\pi}{4}$ since it provides the best match at high-$q$ (low distance from the corner), but it is critical to remember that the tapered edge is not a duplication of the real heterogeneity present at the edge, and hence, some mismatch between theory and experiment is to be expected. For additional details on the selection of $\phi_r = \frac{3\pi}{4}$ see Supplementary Fig. 16.

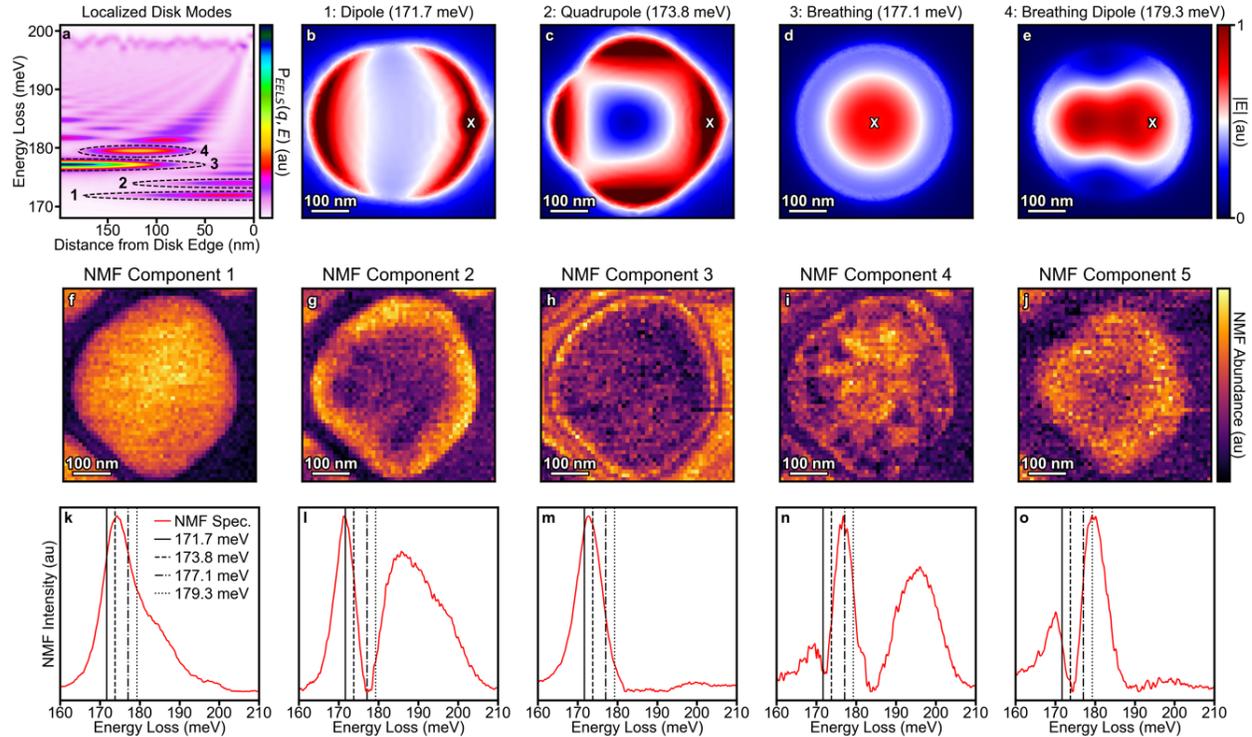

**Supplementary Figure 4. Localized Disk Modes with Non-negative Matrix Factorization.** (a) Simulated EELS for a finite disk of radius 200 nm, showing peaks associated with localized modes. The four most dominant localized modes are highlighted, and their field profiles are simulated and shown in (b-e), with the simulated probe position shown with the white cross. Mode 1 is a dipole-mode at 171.7 meV (b), mode 2 is a quadrupole mode at 173.8 meV (c), mode 3 is a breathing mode at 177.1 meV (d), and mode 4 is a breathing dipole mode at 179.3 meV (e). Non-negative matrix factorization (NMF) is a unmixing technique that is used to separate localized modes in visible plasmonics[2] and we apply it to this dataset. We perform a 5 component NMF decomposition which creates spatial abundance maps (f-j) and corresponding spectral endmembers (k-o).

There are significant differences between the NMF components and the localized modes but recall there is more than just the localized polaritonic response of the disk modes induced by the lacey carbon in the EELS signal. The dataset also contains the infinite slab excitations that do not interact with the lacey carbon, as well as the direct excitation of the TO and LO phonons. As a result, much of the high energy features in the spectral end members do not directly correspond to features in the simulations. Also, the simulations are for a single probe position corresponding to only one polarization, the real experiment over many probe positions excites all polarizations so all modes appear as rotationally symmetric in the corresponding energy-filtered maps. However, both spatially and spectrally the dominant peaks in the NMF components match the simulated disk modes. NMF component 2 is peaked directly at the dipole mode energy and is localized around the outer edge. For NMF component 3, the peak energy is just below the quadrupole energy, but at a higher energy than the dipole, which is consistent with theory. More encouraging is the localizations of the modes. NMF component 3 is tightly bound to the hole edge, while component 2 extends further in. Comparing those localizations to the profiles in in (b) and (c) shows that the dipole mode is also more delocalized here with respect to the quadrupole mode, corroborating our assertion that NMF component 3 represents the quadrupole mode.

The peak energy in NMF components 4 matches the breathing mode peak, and the peak spatial intensity is in the center of the hole not around the edges. This component has clearly incorporated much of the LO phonon excitation, as is evidenced by the high energy peak, which explains why the contrast is more uniform in the abundance map than the other NMF components. Lastly, NMF component 5 shows a pleasing match both spatially and spectrally to the dipole breathing mode, exhibiting a clear annular intensity contrast corresponding to all the different polarizations that would be excited at the different probe positions. NMF component 1 does not strongly represent any particular polariton mode, so likely this is the component that contains the signal from the infinite slab excitations.

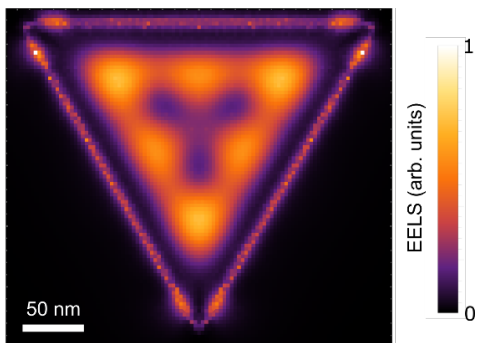

**Supplementary Figure 5. Localized Modes at the Corners of hBN Nanotriangles.** Numerically calculated energy-filtered EELS map (at energy loss of 190 meV) for a beam scanned across a finite equilateral hBN triangle. The nanostructure is 12 nm thick. The numerically-calculated energy-filtered map corresponds to localized modes emerging from edge-confined HPhPs yielding high loss probability at the triangle edges, which is further enhanced very close to the triangle tips. Similar behavior is observed also in the experimental analysis of a real hBN corner and the numerically calculated loss probability for a semi-infinite hBN corner at 180 meV in Main Text Figs. 5b and 5c, respectively and in Supplementary Figure 6b and 6f.

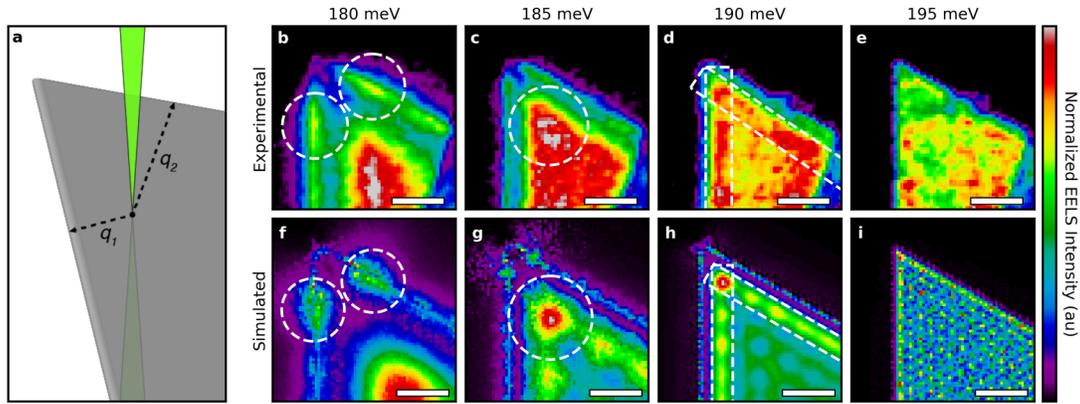

**Supplementary Figure 6. Highlighted Similarities in Simulated and Measured Interference Effects in hBN Corner.** Here, we show a figure equivalent to Main Text Figure 5 where we individually highlight the similar effects observed in EELS and the simulations for the interference patterns in a hBN corner. In (b) and (f), the 180 meV window, we highlight the localized edge polariton mode shown in Supplementary Figure 5. In (c) and (g), the 185 meV window, we highlight a spot near the tip of the of the corner, where reflections from both edges positively interfere to create a hotspot. In (d) and (h), the 190 meV window we highlight the observation of linear positive interference orders that can be seen on close to each flake edge. It is important to note that the experimental images have been denoised here (not in the Main Text) with a Gaussian blur with **σ**=0.75 pixels to help emphasize these features. However, especially in the (e), the 195 meV window, the Gaussian Blur can often emphasize noise, demonstrating that the Gaussian blur denoising process is prone to artifacts. However, the features highlighted here are sufficiently symmetrical and high enough in intensity for us to believe that they are genuine features.

## Supplementary Discussion 1: Fitting of Polariton Peaks in EELS

Due to the densely packed nature of the hyperbolic phonon polariton (HPhP) peaks, traditional spectral fitting with Lorentzians and Gaussians was not found to be the most accurate method of displaying the peak energies at differing probe positions in the EELS line profiles. Instead, we use a Python local peak fitting routine in the sci-kit image library, *peak_local_max*, to identify all peaks in the image, then manually identify the starting position of the peak we wish to track along the line profile and cut out all peaks that are not the closest to the previous peak. This process, which we outline in Supplementary Fig. 7, allows effective tracking of dispersive peaks in datasets when there are multiple peaks present. We also use Savitky-Golay filtering along the spectral axis to denoise the datasets and binning across neighboring probe positions (distances from the edge) to improve the quality of the fit. The following figures (Supplementary Figs. 8 and 9) serve as both as additional demonstration of the process and validation of the fitting process.

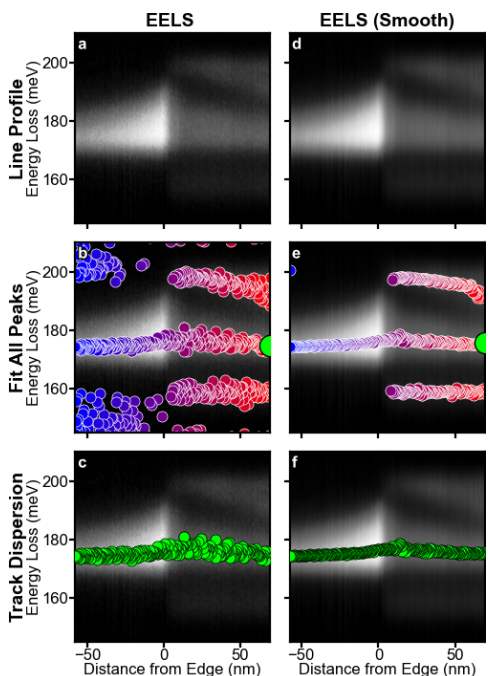

**Supplementary Figure 7. Fitting Peaks in the Line Profile from Main Text Fig. 1.** We take the full dataset and use the *peak_local_max* function to find all peaks in the dataset. This is done as a series of 1D peak finding so the colors (red-blue). Each of them represents a different peak finding irrespective of the previous. We then select the peak to be tracked (green) at one probe position, then move all the way through the dataset accepting only the peak closest the peak in the previous probe position to be accepted as the tracked peak. The result is the green circles shown in the final panel. Panels (a-c) show the as-acquired dataset. Panels (d-f) show the filtered and binned fitting (with significantly improved quality of fit).

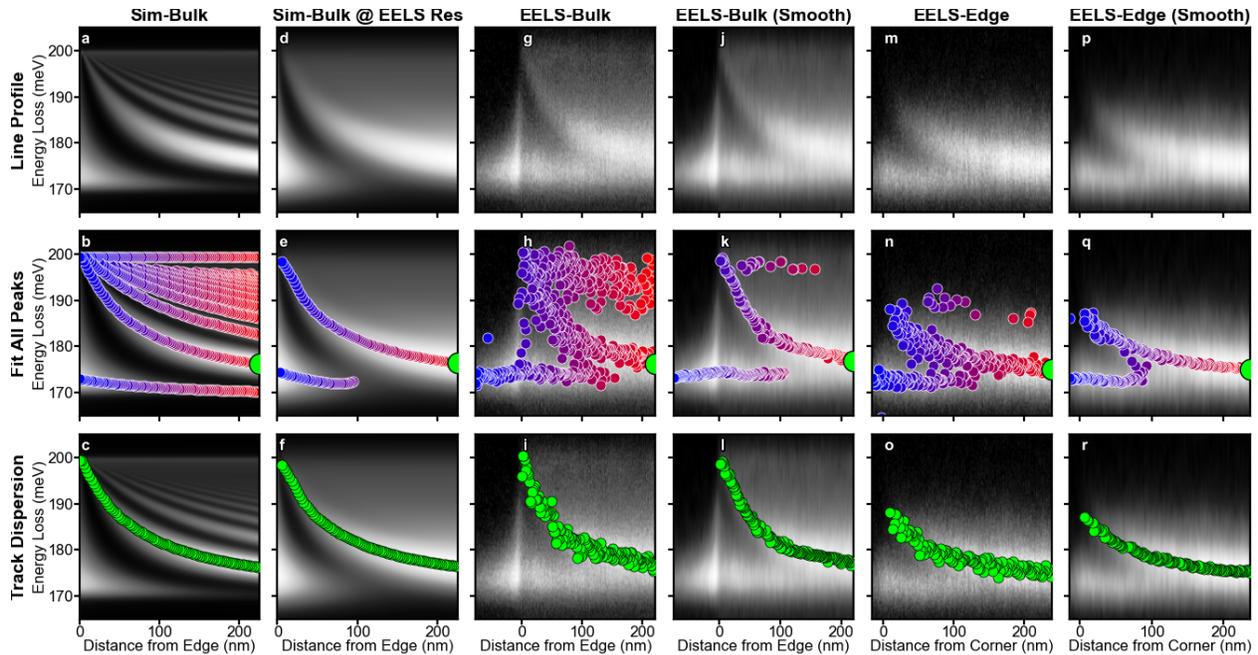

**Supplementary Figure 8. Fitting Peaks in the Dispersion Line Profiles from Main Text Fig. 2 and 3.** The peak-fitting method utilized above is applied to the dispersion datasets shown in Main Text Fig. 2 and Fig. 3, as well as the simulated bulk polariton dispersions. (a-c) Bulk HPhP simulation (without broadening). (d-f) Bulk HPhP simumulation, broadened with the EELS energy resolution. (g-i) As-acquired EELS bulk HPhP line profile, (j-l) Smoothed and binned EELS bulk HPhP line profile, (m-o) as-acquired EELS edge HPhP line profile, (p-r) smoothed and binned EELS edge HPhP line profile.

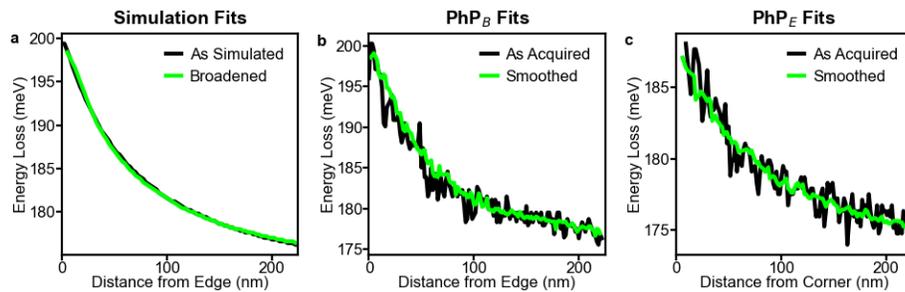

**Supplementary Figure 9. Comparison of Smoothed/Broadened Fits.** (a) Comparison of simulated profiles, showing that no significant change in the measured peak energy of the first order reflected is induced by broadening, confirming that EELS can accurately measure these values. (b) Comparison of bulk polariton fits, showing smoothing/binning does not affect the measured dispersion. (c) Comparison of edge polariton fits, showing smoothing/binning does not affect the measured dispersion. We note here that the fit does not extend all the way to zero. This is due to the suppressed intensity at the corner from destructive interference discussed in Main Text Fig. 5.

# Supplementary Discussion 2: Parameters for Dispersion Measurement

Dataset Calibration

At the high spectrometer dispersions used in this work required for EELS measurements in the infrared, the dispersion can routinely have errors of ~1%, when calibrated by drift tube, due to instabilities in the instrument, necessitating calibration to a known spectral feature. Due to the complex nature of the hBN polariton response the LO phonon is the only logical choice for calibration as for thick flakes it should dominate the spectrum and its energy should not be influenced by localized polariton modes or thickness. Supplementary Fig. 10 shows the calibration of the datasets used in the manuscript. In Supplementary Figs. 10a and 10b we show simulations for the infinite slab response of hBN flakes at ascending thicknesses, both in the actual energy resolution (Supplementary Fig. 10a) and broadened with a 6 meV wide Gaussian to match the EELS energy resolution (Supplementary Fig. 10b). The simulations show that, while the polariton peak energy is strongly influenced by thickness the LO phonon, it only increases in intensity at higher and higher thicknesses.

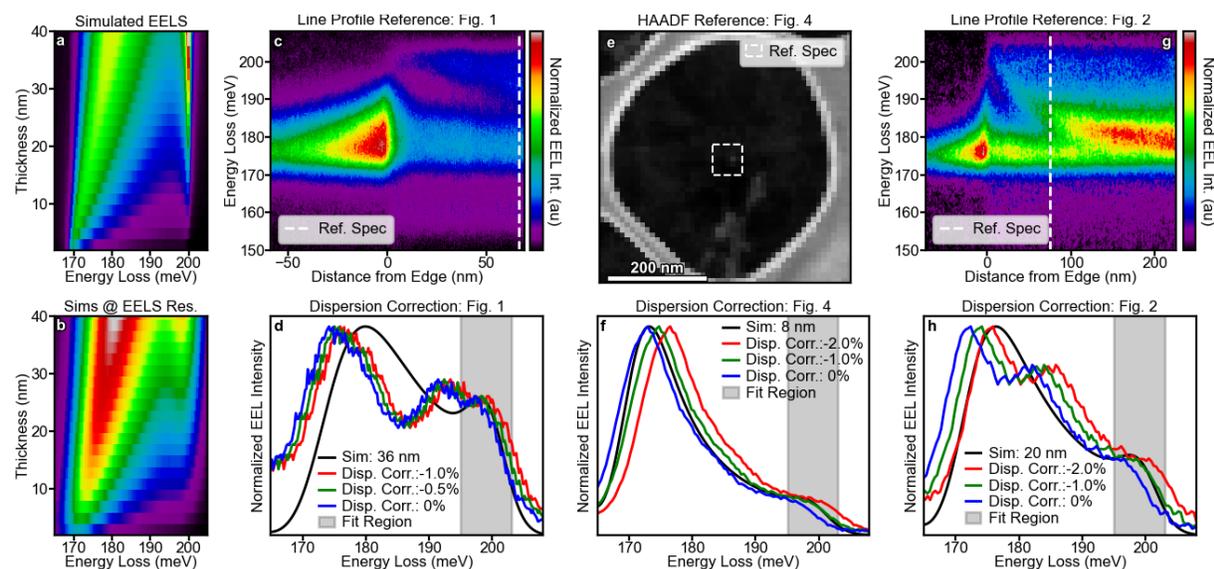

**Supplementary Figure 10. Calibrating Spectrometer Dispersion to LO phonon peak.** (a,b) Simulated EELS spectra at different thicknesses, showing that the LO peak energy does not change as a function of thickness. (c) Calibration of the line profile in Main Text Fig. 1. The white dashed line is the reference point used for the calibration spectrum. (d) Comparison of the LO phonon shows the best match to the LO peak at -0.5% dispersion correction. (e) Calibration of the hole in lacey carbon spectrum image shown in Main Text Fig. 4, using the reference point from the center of the hole. (f) Best match to LO phonon at -1.0% dispersion correction. (g) Calibration of the line profile in Main Text Fig. 2. The white dashed line is the reference point used for the calibration spectrum. (h) Comparison of the LO phonon shows the best match to the LO peak at -1.0% dispersion correction. Datasets in Main Text Figs. 2-5 in the main text all use the same dispersion from the same day. The agreement between (f) and (h) allows us to apply the -1.0% dispersion correction to the Main Text Figs. 3 and 5 datasets.

The calibration for the line profile in Main Text Fig. 1 is shown in Supplementary Figs. 10c and 10d. Here, we take the probe position furthest from the edge for calibration to avoid signal mixing with the dispersive interferometric peak discussed in Main Text Fig. 2 (which can be clearly seen in the line profile shown here in Supplementary Fig. 10c). The spectrum from this probe position is shown in Supplementary Fig. 10d, along with the EELS simulation for a 36 nm thick slab. We carefully use a similar process to determine the effective flake thickness in the next session. Here, the 36 nm thickness is selected to get the best match to the LO peak intensity in the normalized spectrum, which has been artificially increased due to its close proximity to the interference peak (at around 192 meV). Regardless of the proximity of the interference peak, a separate peak can be made out for the LO phonon, which shows the best match to the simulation at dispersion of -0.5% of the as-acquired value (where 0% is the as-calibrated dispersion).

The same process is used for the lacey carbon hole dataset discussed in Main Text Figure 4 in Supplementary Figs. 10e and 10f, where now we use the spectra from the center of the hole as the reference spectrum. Here, we compare to a flake thickness of 8 nm (once again to achieve the best quantitative match to the LO phonon) and see a clear optimum with a -1.0% dispersion correction. Lastly, we apply the process to the bulk line profile from Main Text Fig. 2 in Supplementary Figs. 10g and 10h. Once again, we see a clear optimum at a dispersion correction of -1.0%. The match agreement of the Main Text Figs. 4 and 3 datasets is encouraging since these two datasets were acquired on the same day at the same dispersion so there should be no significant changes. The agreement also allows us to straightforwardly apply the -1.0% dispersion correction to the edge polariton line profile from Main Text Fig. 3 and the corner dataset shown in Main Text Fig. 5 as these were also acquired on the same day at the same dispersion.

Thickness Measurements

It is also critical to ascertain the thickness of the flake as the polaritonic response is highly sensitive to this parameter. Traditional log-ratio thickness measurements can provide accurate estimates but the thickness dependent nature of the polaritonic response enables accurate thickness measurements directly from the spectra. As addressed in Main Text Fig. 4, the detected EELS signal is combination of localized modes interacting with the flake geometry and surrounding heterogeneities with the propagating infinite slab response. We can take a reference where the main HPhP peak is dominated by this infinite slab response and match it to the thickness simulations of the infinite slab response to obtain an accurate thickness measurement for the sample.

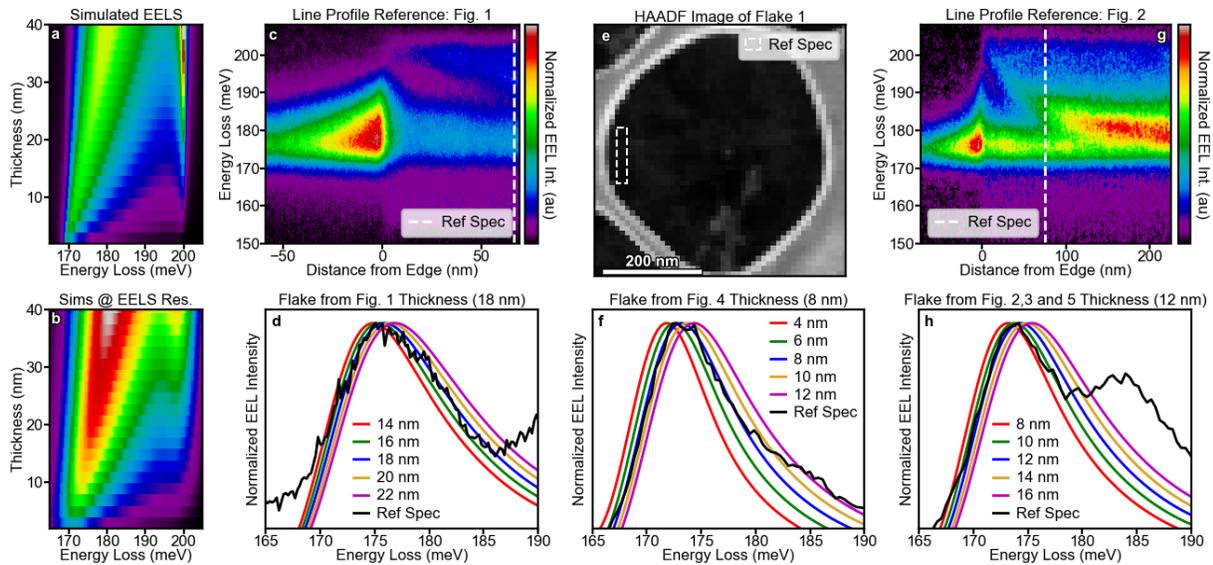

**Supplementary Figure 11. Determining Flake Thickness.** (a,b) Simulated EELS spectra at different thicknesses, showing both the LO peak (not changing in energy) and a dispersive polaritonic peak with energy strongly varying with the slab thickness. (c) Line profile reference from flake in Main Text Fig. 1. The white dashed line is the reference point used for the thickness determination spectrum. (d) Comparison of the reference spectrum with different simulated thicknesses, showing the best match for an 18 nm flake thickness (e) HAADF reference from the flake in Fig. 4. Reference spectra are taken from the edge of the hole. (f) Best match to data at 8 nm flake thickness (g) Line profile reference from the flake in Main Text Figs. 2, 3, and 5. (h) Good matches are achieved for both 10 nm and 12 nm. In Supplementary Fig. 14, shown later, reflection phase analysis is used to determine that 12 nm presents a better match across the whole dispersion.

Supplementary Fig. 11 shows the thickness determination for the three flakes examined in the paper. Supplementary Figs. 11a and 11b once again show the EELS simulations as a function of slab thickness, Supplementary Figs. 11c and 11d show the thickness determination of the flake in Main Text Fig. 1, Supplementary Figs. 11e and 11f address the thickness of the flake in Main Text Fig. 4, and Supplementary Figs. 11g and 11h address the thickness of the flake in Main Text Figs. 2, 3, and 5. For the line profiles from Main Text Figs. 1 and 2 we use the same reference spectrum as was used in the calibration determination, but for the thickness determination of the flake in Main Text Fig. 4 we use a reference spectrum from the edge. We recall that in Main Text Figs. 4h and 4i we saw that the spectra at the edge of the hole are extremely close to the peak energy of the propagating slab, while those to the center have been blueshifted by the breathing mode.

For the thickness determinations in Supplementary Fig. 11c-f, we observe an excellent match between the experimental and simulated peaks at a single thickness (18 nm for the flake in Main Text Fig. 1 and 8 nm for the flake in Main Text Fig. 4). More than that, we see a promising match in the linewidth of the experimental peak in the entire spectral region where the signal is dominated by the propagating mode. The influence of localized/interfering propagating modes can be seen starting at around 185 meV in Supplementary Fig. 11d and 180 meV in Supplementary Fig. 11f.

In Supplementary Fig. 11h, there is not a clear match to a single thickness, on the low-energy/low-q side of the propagating peak we see a better match to 12 nm, while on the high energy/high-q side of the propagating peak a better match observed for 10 nm. From this comparison alone it is not clear which thickness presents the better match to data, however, once we include reflection simulations into the analysis (shown later in Supplementary Fig. 14), we find that a flake thickness 12 nm presents a much better match experiment, and we will choose this value.

Reflection Phase: Bulk Polariton

The reflection phase $\phi_R$ is critical both for the experiment and the theory of the dispersion measurements in polariton interferometry. For the experiment, it is needed as a transformation that allows us to convert the distance to the edge of a flake to a wave vector in a dispersion by means of the equation $q = \frac{2\pi - \phi_R}{2x}$, where we assume that the phase is a real constant. For the theory, the reflection phase has a significant impact on modeling of EELS polariton interferometry line profiles shown in Main Text Figs. 2 and 3.

We use the following approach to theoretically check if the reflection phase is constant and to obtain its value: we simulate the induced electromagnetic field for a certain beam position and then try to fit the resulting interference minima.

In particular, the electromagnetic field generated by a bulk HPhP propagating to an edge and then reflecting back to interfere with itself can be modelled as a superposition of two waves with cylindrical symmetry. The field associated with the polaritons propagating from the beam position is $\mathbf{E}_s \propto \mathbf{E}_{s0} \exp(iqR)/\sqrt{R}$. The reflected polaritonic field $\mathbf{E}_{s'} \propto \mathbf{E}_{s'0} \exp(-iqR' - \phi_R)/\sqrt{R'}$ can be understood as emerging from a virtual source placed outside the edge. We assume that the electric field $\mathbf{E}_{s0}$ and $\mathbf{E}_{s'0}$ is modulated by the phase factor containing the dominant polariton in-plane wave vector

$$q = \frac{2}{t} \mathrm{Re}\left[i\sqrt{\frac{\epsilon_z}{\epsilon_R}} \mathrm{atan}\left(\frac{i}{\sqrt{\epsilon_R \epsilon_z}}\right)\right], (1)$$

where $t$ is the flake thickness, and the dielectric parameters are defined for hBN in Supplementary Table S1, which is shown in Supplementary Discussion 3, along with the derivation of this analytical form. We futher define $R = \sqrt{(x - x_b)^2 + y^2}$ and $R' = \sqrt{(x + x_b)^2 + y^2}$, where $x, y$ are spatial coordinates perpendicular to the beam trajectory, $x_b$ is the beam distance from the edge at $x = 0$. The intensity amplitude of the resulting field $|\mathbf{E}_s + \mathbf{E}_{s'}|^2$ is modulated by an interference term proportional to

$$I_{IF} \propto I_0 + \cos(q(R + R') + \phi_R),$$

where $I_0$ is an intensity offset. The above formula for the intensity yields interference minima when

$$q(R + R') + \phi_R = (2m + 1)\pi, (2)$$

where $m$ is an integer. It can be shown that the resulting minima lay on hyperbolas or ellipses in the $xy$ plane. Supplementary Fig. 12 shows the analytically calculated positions of interference minima from Supplementary Eq. (2) (dashed lines of different colors corresponding to varying $m$) overlaid on numerically calculated plots of electrostatic potential (resembling the out-of-plane electric field component $E_z$) for two distinct energies. The best fit between the analytically and numerically calculated minima in the field patters is obtained for reflection phase close to $\pi/2$ which, as we checked, hold also for other energies and various slab thicknesses below $\sim 50$ nm.

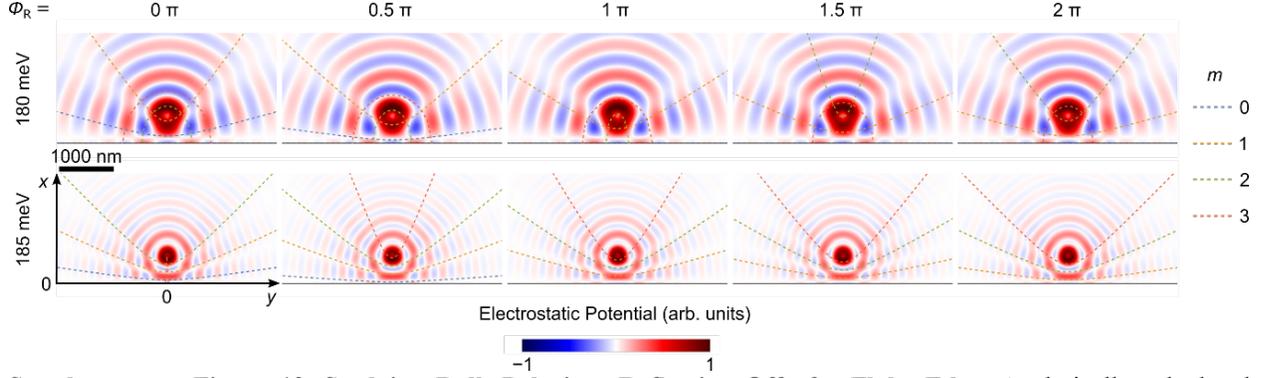

**Supplementary Figure 12. Studying Bulk Polariton Reflection Off of a Flake Edge.** Analytically calculated positions of minima according to Supplementary Eq. (2) resulting from interference of bulk phonon polaritons freely propagating and those reflected off the edge boundary (colored dashed lines) on top of the numerically calculated electrostatic potential plotted right above the upper surface of a 18-nm thick hBN slab (the $xy$ plane). The slab edge is situated at $x = 0$ (denoted by black lines), while the electron beam is passing perpendicularly to the screen and crossing the slab at position $x_b = 500$ nm, $y_b = 0$ nm. The comparison is shown for energies of 180 meV (upper row) and 185 meV (lower row) for varying reflection phase in the analytical calculation as denoted above each column.

To model EELS spectra taking into account interference of HPhPs, we have to evaluate the resulting electric field at the electron beam position $x_b, y_b = 0$. Based on above arguments, we use the following analytical phenomenological model for the resulting EELS probability $\Gamma_{\text{trunc}}$:

$$\Gamma_{\text{trunc}}(\omega, x_b) \approx \int_0^{Q_c} dQ \; P(Q, \omega) \left[1 + \cos(2Q|x_b| + \phi_R)\right], \quad (3)$$

where $P = P_{\text{bulk}} + P_{\text{guid+Begr}}$ which are defined implicitly within Supplementary Eqs. (14) and (15) presented in Supplementary Discussion 3. However, we note that if the reflection phase is imaginary, there will be a reduction of the interference term and more dominant contribution of the freely propagating polaritons.

Reflection Phase: Edge Polariton
For the edge-guided HPhP, we use the following approximate form for the analytical dispersion

$$q_y = \frac{\pi}{t} \text{Re} \left[ \frac{1}{\epsilon_z} \frac{\epsilon_R \epsilon_z - 1}{1 - \epsilon_R^2} \right], \quad (4)$$

which is discussed later in Supplementary Discussion 3 and reproduces reasonably well numerically calculated results for an exact 3D geometry of hBN truncation with the flat edge (see Main Text Fig. 3f in the main text). Importantly, by comparing the dispersion relations of edge-guided polaritons in Supplementary Eq. (4) to bulk-guided polaritons in Supplementary Eq. (1), we see that the former dispersion exhibits higher $q$-vectors at a fixed energy. Hence, wavelength of edge-polaritons is shorter compared to the bulk polaritons (as can be seen by comparing the upper limits of the EELS excitations in Main Text Fig. 2 and Fig. 3 in the main text), which has also been shown with scanning near-field optical microscopy[3]. For the same reasons, the edge polaritons are more confined.

The EELS spectra, taking into account the possible interference of freely propagating edge HPhPs and those reflected from the edge corner, are modeled following a scheme similar to that in Supplementary Eq. (3):

$$\Gamma_{\text{corner}}(\omega, y_b) \approx \int_0^{q_{y,c}} dq_y \; P(q_y, \omega) \left[1 + \cos(2q_y|y_b| + \phi_R)\right], \quad (5)$$

where $|y_b|$ is the beam distance from the corner, $\phi_R$ is now the reflection phase at the hBN corner, $P(q_y, \omega)$ is the $q_y$-resolved EELS probability at the sample edge, and $q_{y,c}$ is the corresponding wave vector cutoff. $P(q_y, \omega)$ is calculated numerically (as later described in the last section of the Supplementary Discussion 3) and is shown in Main Text Figs. 3f and 3g, in the main text for a straight and tapered edge geometry, respectively.

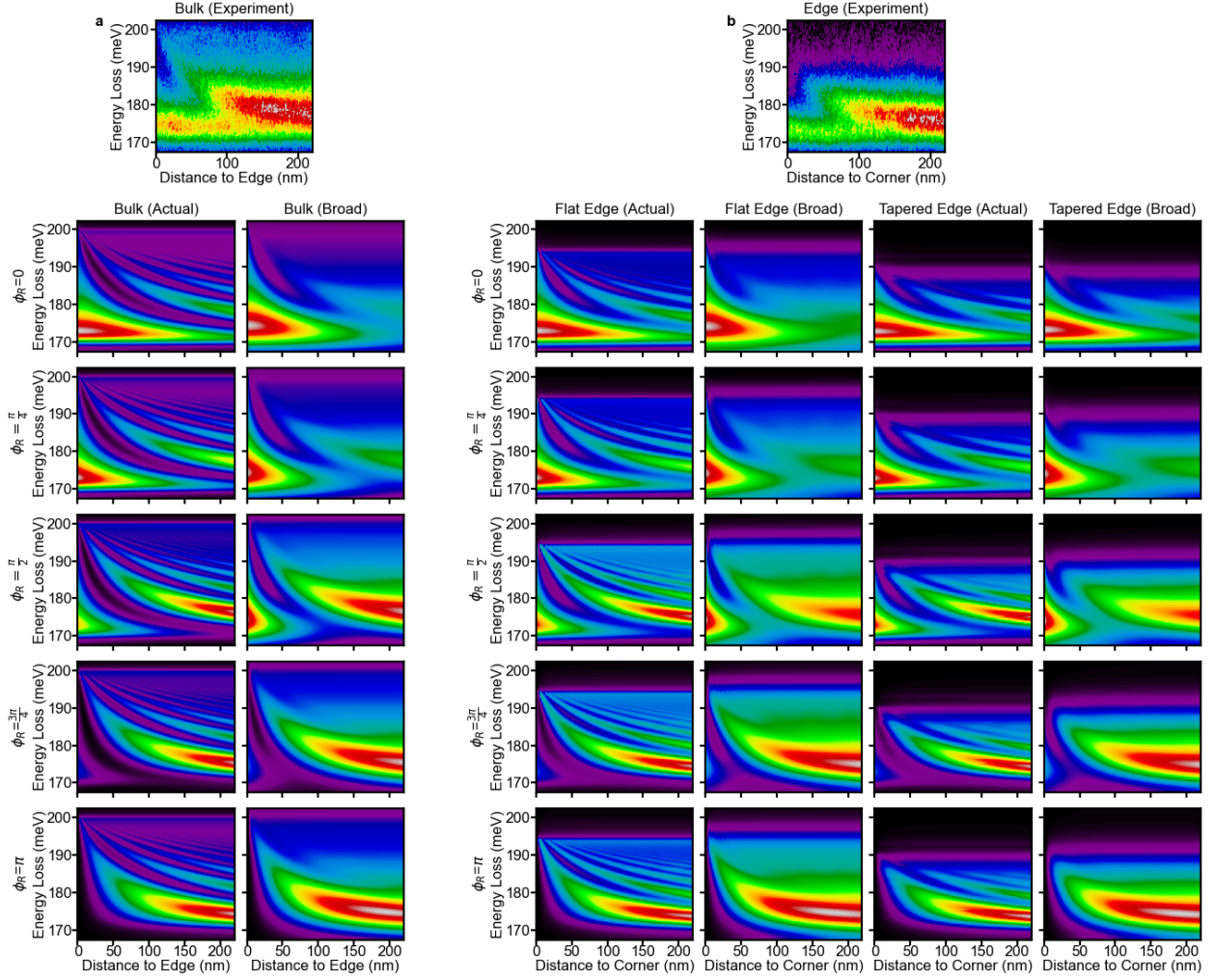

**Supplementary Figure S13. Analytically Calculated EELS Probability as a Function of Beam Position.** (a) Bulk polariton line profiles from the experiment and analytically calculated reflection of the bulk polariton from the edge calculated according to Supplementary Eq. (3) at reflection phases, $\phi_R = 0, \frac{\pi}{4}, \frac{\pi}{2}, \frac{3\pi}{4}, \pi$. (b) Edge polariton line profiles from the experiment and both flat/tapered edge simulations semi-analytically calculated EELS probability Supplementary Eq. (5) as a function of beam distance from the corner at the different reflection phases.

Reflection Phase Determination: Theory vs. Experiment

Supplementary Fig. 13 shows analytically calculated loss probabilities for varying beam positions and different reflection phases. In Supplementary Fig. S13a, we consider reflection of the bulk polaritons, as determined by Supplementary Eq. (3). The semi-analytically calculated loss probabilities for flat and tapered edge line profiles are shown in Supplementary Fig. S13b, as determined by Supplementary Eq. (5). In the actual, as-calculated line profiles many higher orders of reflection (including the $0^{th}$ order dominant at the edge of the flake) are identifiable, however in the spectra that have been broadened to match the EELS resolution only the $0^{th}$ order and $1^{st}$ order peaks are visible. The simulated intensity distributions in Supplementary Fig. 13 show that for $\phi_R < \pi/2$ the intensity in the 176 meV range should decrease significantly as the probe moves away from the edge. Such intensity decreases are not observed in experiment, making the data qualitatively inconsistent with the lower values of $\phi_R$.

With these simulations we can return to the partly ambiguous result shown in Supplementary Fig. 11, where the data matched to thicknesses of both 10 nm and 12 nm. We can perform a similar comparison to the different reflection phase values (as done in Supplementary Figs. 2 and 3) for both 10 nm and 12 nm thicknesses, which is shown in Supplementary Figure 14. The bulk polariton experimental line profile is shown in Supplementary Fig. 14a, with the peak fits, and the comparison to 10 nm reflection phases is shown in Supplementary Fig. 14b and the 12 nm reflection phases is shown in Supplementary Fig. 14c. Even here, we see comparable agreement to the 10 nm with $\phi_R = \frac{\pi}{4}$ to 12 nm with $\phi_R = \frac{\pi}{2}$. However, as can be observed in the simulations for a 10 nm with $\phi_R = \frac{\pi}{4}$ (Supplementary Fig. 14d) and a 12 nm flake with with $\phi_R = \frac{\pi}{2}$ (Supplementary Fig. 14e). The relative intensity of the $0^{th}$ order and $1^{st}$ order peaks for $\phi_R = \frac{\pi}{4}$ do not match what is observed in experiment.

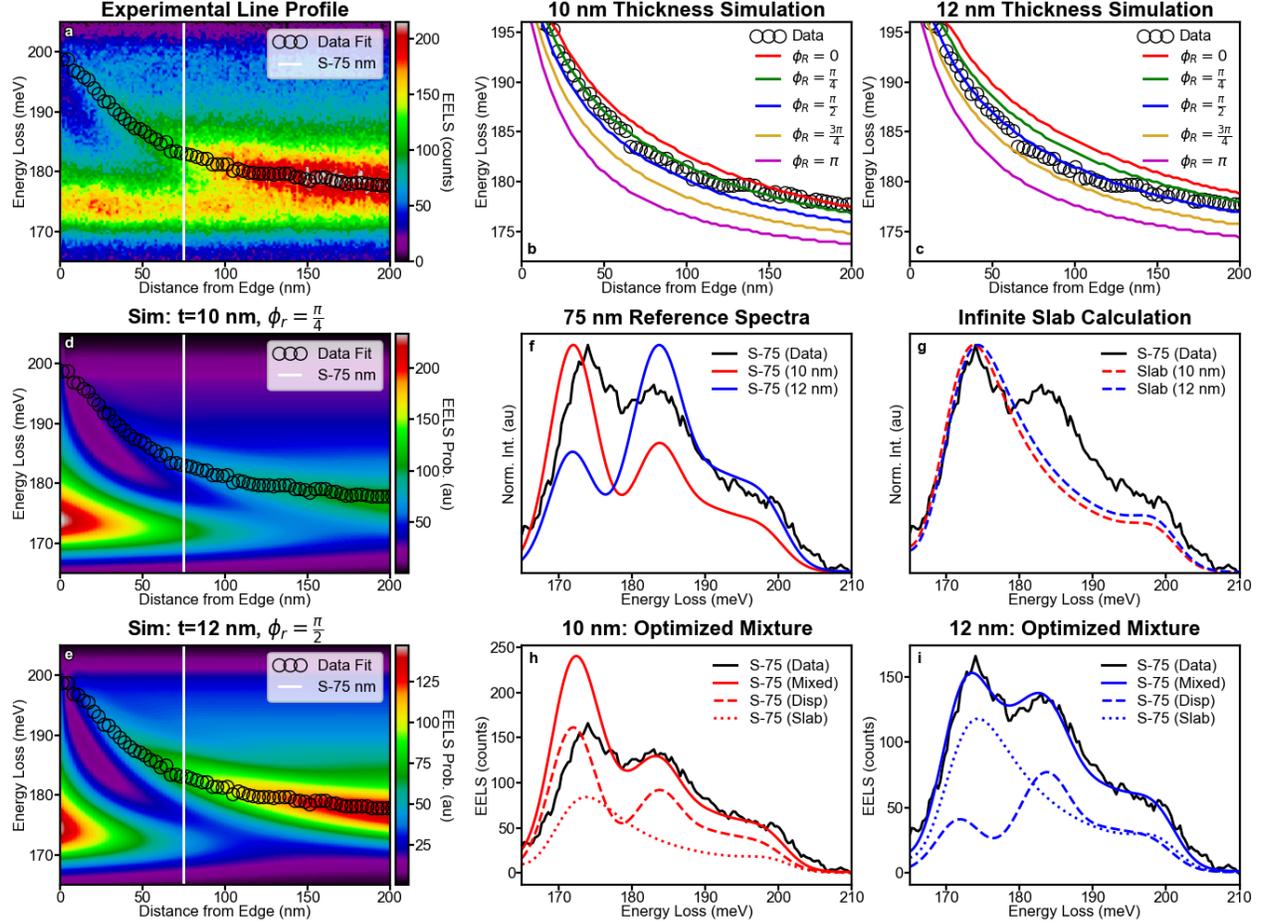

**Supplementary Figure 14. Comparison of $\phi_R = \frac{\pi}{4}$ and $\phi_R = \frac{\pi}{2}$ Simulations** (a) Bulk polariton line profiles from the experiment. (b) Comparison of experiment to simulated line profiles at different reflection phases for a 10 nm thick flake, showing the best match at $\phi_R = \frac{\pi}{4}$. (c) Comparison of experiment to simulated line profiles at different reflection phases for a 12 nm thick flake, showing the best match at $\phi_R = \frac{\pi}{2}$. (d) Simulated line profile for a 10 nm thick flake with $\phi_R = \frac{\pi}{4}$. (e) Simulated line profile for a 12 nm thick flake with $\phi_R = \frac{\pi}{2}$. (f) EELS Spectra at 75 nm from the flake edge for experimental and simulated spectra. (g) Propagating slab EELS simulations for both thicknesses. (h) Optimal combination of S-75 and Slab calculation for a 10 nm thick flake, showing quantitative and qualitative mismatch. (i) Optimal combination of S-75 and slab calculations for a 12 nm thick flake, showing quantitative and qualitative agreement.

To further expand on this, we examine the spectra from a probe position 75 nm away from the flake edge (S-75) from all three datasets (plotted in Supplementary Fig. 14f). Here, it is tempting to say that the 10 nm spectrum agrees better with experiment than the 12 nm spectrum, but one must keep in mind that the reflection phase might be complex and that a higher portion of freely propagating polaritons not contributing to the interference can participate in the measured spectra. As seen in Fig. 4 of the main text, in order to accurately model the real EELS signal, we thus combine the propagating slab excitations with the localized/interferometric modes. The propagating slab excitations for a 10 nm flake and 12 nm flake are shown in Supplementary Fig. 14g and look nearly identical except for the small blue shift due to the increased thickness observed in the 12 nm flake. However, when combined with the S-75 mode, it becomes clear that no combination can allow the $\phi_R = \frac{\pi}{4}$ to quantitatively reproduce the experimental spectrum. As shown in Supplementary Fig. 14h, if the combined 0$^{th}$ order and propagating peak is quantitatively correct than the 1$^{st}$ order reflection mode and LO phonon are too small, and if the 1$^{st}$ order reflection and LO modes are correct than the combined propagating slab and 0$^{th}$ order peak is too high. Moreover, due to the reduced relative intensity of the 0$^{th}$ order mode with respect to the 1$^{st}$ order mode in the $\phi_R = \frac{\pi}{2}$ calculation, no such tradeoff is necessary for the 12 nm thickness. Supplementary Fig. 14i shows that a combination of the propagating slab and interferometric modes exists where the experimental spectrum is reproduced qualitatively and quantitatively. This comparison allows us to select 12 nm for the flake thickness discussed in Supplementary Fig. 11 and verify that $\phi_R = \frac{\pi}{2}$ is the correct selection for the bulk polariton dispersion transformation/simulation.

Reflection Phase & Spatial Calibration
In addition to the reflection phase, it is critical to rigorously define the flake edge in the hyperspectral dataset. In a theoretical model, the edge of the flake is well defined, in an experiment this is not the case, especially for the step-like edges found in most 2D materials that has an in-plane width of several nm. In order to perform the spatial calibration, we perform a series of analyses such as those found in Supplementary Figs. 2 and 3 where we compare the fits of the experimental line profile to the fits of the simulated line profile, however here we alter the x-axis for the experimental dataset by changing the pixel in the line profile labeled as the edge or corner ($x=0$).

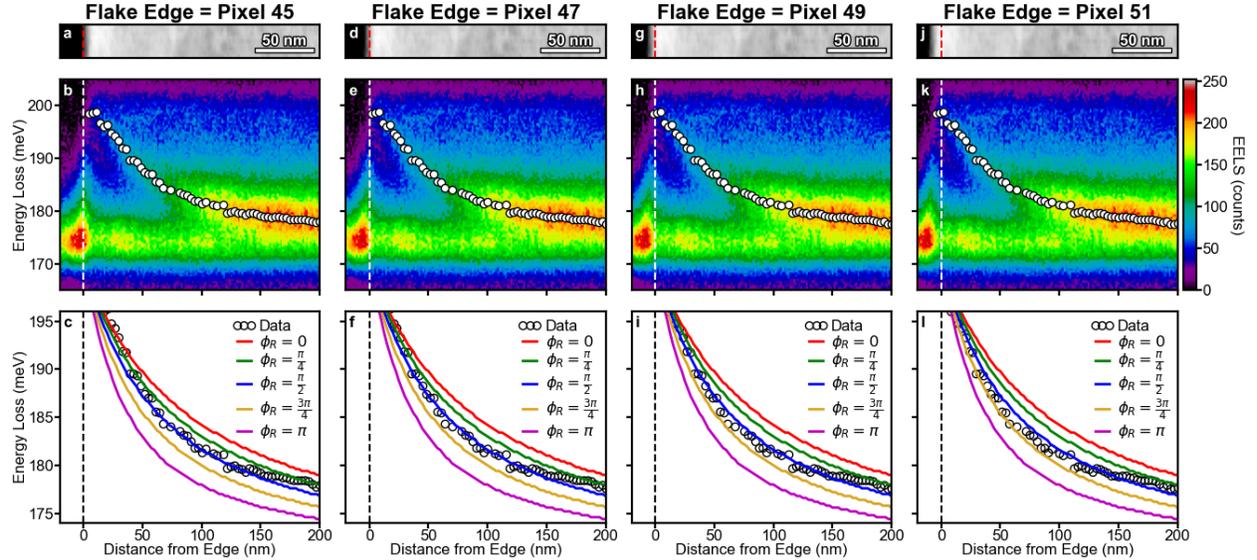

**Supplementary Figure 15. Comparison of Bulk Polariton Reflection Phases and Spatial Calibrations.** (a-c) Comparison of experiment simulations for the bulk polariton line profile, where the experimental line profile is calibrated such that the flake starts at pixel 45 in the horizontal direction. (a) HAADF reference, (b) line profile and experiment fit, (c) data and simulation at all five reflection phase fits. Dashed lines correspond to pixel 0 in all three images. (d-f) Calibration for flake edge at pixel 47. (g-i) Calibration for flake edge at pixel 49. (j-l) Calibration for flake edge at pixel 51. We find the best match for flake edge at pixel 49 and $\phi_R = \frac{\pi}{2}$.

Supplementary Figure 15 shows the flake edge comparisons for $x=0$ calibrations on the bulk line profile experiments. The top panel of all four calibrations shows the calibrated HAADF reference, here we can see that Flake Edge = Pixel 45 corresponds to the edge of the flake being counted as the outermost point of the edge of the flake, while Pixel 49/51 correspond to the start of the heterogeneous flake edge. For all these varying calibrations, it can be observed that the best match is achieved for flake edge = pixel 49 with $\phi_r = \frac{\pi}{2}$, which is in good agreement with the prediction from theory that suggests that $\phi_R = \frac{\pi}{2}$ is the optimal reflection phase to use for these experiments.

The same analysis is performed for the experimental edge polariton line profile with the tapered edge simulations in Supplementary Fig. 16. The dataset used for the experiment is a 2D spectrum image, the corresponding HAADF image of which is shown in the top panel for each calibration in Supplementary Fig. 16. Only the pixels highlighted in cyan in the image are used for the line profile. We experimented with different numbers of binned pixels and did not find that this changed the measured dispersion in any significant way, but that the best combination of signal-to-noise ratio (while staying outside the flake) was to take the two pixels just outside of the flake.

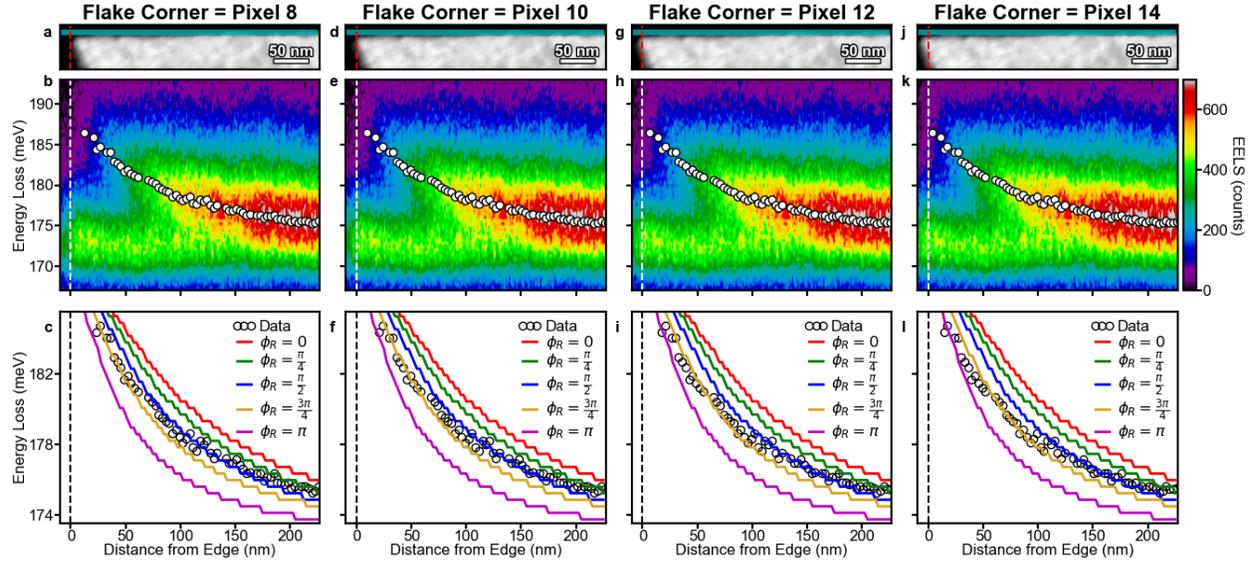

**Supplementary Figure 16. Comparison of Edge Polariton Reflection Phases and Spatial Calibrations.** (a-c) Comparison of the experiment and tapered edge simulations for the edge polariton line profile, where the experimental line profile is calibrated such that the flake corner starts at pixel 8 in the horizontal direction. (a) HAADF reference, (b) line profile and experiment fit, (c) data and simulation at all five reflection phase fits. Dashed lines correspond to pixel 0 in all three images. (d-f) Flake corner = pixel 10 calibration. (g-i) Flake corner = pixel 12 calibration. (j-l) Flake corner = pixel 14 calibration. Best match: flake edge = pixel 10 and $\phi_R = \frac{3\pi}{4}$; however, a strong match is not achieved, likely due to fundamental differences between the heterogeneous edge in the real flake and the modeled 45° tapered edge, as well as due to influence of the actual corner shape.

With the edge polaritons, there is not a single simulation that exhibits agreement as strong as the agreement observed for the bulk polariton in Supplementary Fig. 15, this is likely due to the fact that the edge polariton is much more strongly altered by the presence of heterogeneity at the edge of the flake than the bulk polariton. The actual shape of the flake (i.e. the sharp corner geometry) can also alter the results. In Supplementary Fig. 16, we determine the best match to data to be Flake Corner = Pixel 10 with $\phi_R = \frac{3\pi}{4}$ due to the match at the low-distance (high-q) limit.

While we can say that $\phi_R = \frac{3\pi}{4}$ provides the best match to the simulated tapered edge dispersion, the preceding validation all uniformly suggest that the correct reflection phase to use in interferometry experiments is $\phi_R = \frac{\pi}{2}$. The influence of the tapered edge on the resulting total field is shown in Supplementary Fig. 17. Although the field pattern inside the slab containing contribution of higher-order HPhPs is changed when comparing the straight and the tapered edge, the field outside the slab dominated by the fundamental bulk polariton mode and exhibits minima and maxima at nearly identical positions from the edge boundary at $x = 0$. Additionally, the field profile at the edge of flake is significantly modified due to the edge tapering, demonstrating that the electrodynamics at the edge is fundamentally modified by the presence of heterogeneity. The distinction could potentially explain why the bulk polariton dispersion matches the expected $\phi_R = \frac{\pi}{2}$, while the edge polariton dispersion does not.

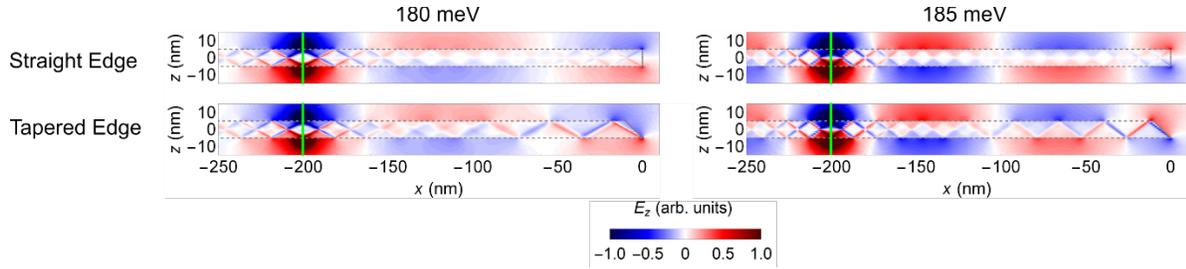

**Supplementary Figure 17. Comparison of Electric Field Excited at the Straight Edge and the Tapered Edge.** We plot the total electric field component $E_z$ for the electron beam passing at the distance $x_b = 200$ nm from the straight edge (upper row) and tapered edge (lower row). We consider loss energies of 180 meV (left column) and 185 meV (right column). hBN boundaries are plotted by dashed gray lines, showing a slab of 10 nm thickness with either straight or tapered edge truncation. The positions of extremities and zeros of the electric field outside the slab are very close for both truncation geometries, implying that the reflection phase of bulk polaritons in both cases is similar.

As a final note on the calibration, we consider the effect of small miscalibrations in the scan coils of the microscope. Such miscalibrations often result in errors of 2-3% in the pixel size of STEM acquisitions. Scan coil miscalibrations have a much less drastic effect on the polariton interferometry than miscalibrations of the spectrometer dispersion. As demonstrated in Supplementary Fig. 18, even for changes of ±5% in the scan calibration, no significant changes in the probe position dependence of the peak or the transformed dispersion are observed. As a result, we use the nominal value for the scan calibration in all datasets.

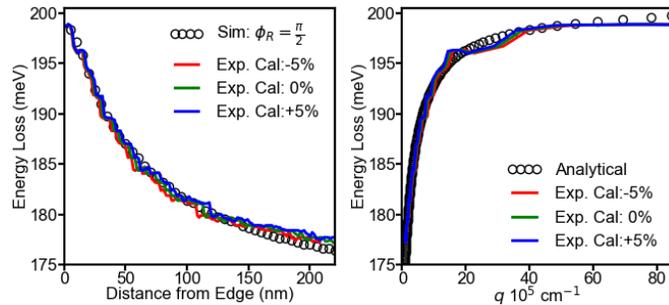

**Supplementary Figure 18. Scan Calibration for Dataset.** Comparison of the scan calibration for the position dependent first-order peak position and transformed dispersion for ±5% errors.

# Supplementary Discussion 3: Methodology and Background for EELS Theory

Dielectric Response and Electromagnetic Waves in Bulk hBN

Hexagonal boron nitride (hBN) is composed of individual layers in which B and N atoms are strongly bound forming a honeycomb lattice in the $xy$ plane. These layers are stacked along the $z$ direction, typically following an AB pattern, and held together by weak van der Waals forces. Such in- and out-of-plane anisotropy in strength of the bonds forming the lattice gives rise to a strongly anisotropic vibrational response of hBN, which in bulk hBN (in practice containing more than 10 layers[4]) can be described by a diagonal dielectric tensor with two distinct components for the in-plane and out-of-plane response: $\hat{\epsilon}_{hBN} = \text{diag}(\epsilon_R, \epsilon_R, \epsilon_z)$. The spectral dependence of each of the individual components can be described by using the Lorentz-Drude model as

$$\epsilon_{R/z} = \epsilon_{\infty,R/z} \left(1 + \frac{\omega_{LO,R/z}^2 - \omega_{TO,R/z}^2}{\omega_{TO,R/z}^2 - \omega^2 - i\gamma_{R/z}\omega}\right) \quad (6)$$

with angular frequency $\omega$, TO/LO frequencies $\omega_{TO/LO}$, high-frequency response $\epsilon_\infty$, and damping $\gamma$ corresponding to either $R$ or $z$ component. In Supplementary Table 1, we summarize the values for these parameters adapted for hBN[5].

Interestingly, hBN exhibits two Reststrahlen bands (RBs) between the TO and LO frequencies when either $\epsilon_R$ or $\epsilon_z$ is negative. In a conventional isotropic material, the appearance of the RBs indicates that propagation of electromagnetic waves in the bulk material is forbidden in this restricted energy range. However, the situation in hBN is more complex due to its anisotropy. The wave vector components $\mathbf{q} = (q_x, q_y, q_z)$ of the (extraordinary) electromagnetic waves propagating in bulk hBN need to fulfill the dispersion relationship:

$$\frac{q_z^2}{\epsilon_R} + \frac{q^2}{\epsilon_z} = \frac{\omega^2}{c^2}, \quad (7)$$

where we define $q^2 = q_x^2 + q_y^2$ and $c$ is the speed of light in vacuum. Upon inspection of Supplementary Eq. (7), we find that the isofrequency surfaces $\omega(\mathbf{q}) = $ constant form hyperboloids of either type I in the lower RB or type II in the upper RB. The hyperbolic isofrequency surfaces determine the peculiar propagation of waves in bulk hBN and are responsible for the different nature of polaritonic excitations in hBN compared to conventional isotropic materials. Waves propagating in bulk hBN can be efficiently excited by fast electrons[6], which would be reflected in electron energy-loss spectra for a beam passing through a very thick hBN film.

| $\epsilon_{\infty,R}$ | 4.9 | $\epsilon_{\infty,z}$ | 2.95 |
|---|---|---|---|
| $\omega_{TO,R}$ | 1360 cm$^{-1}$/168.6 meV | $\omega_{TO,z}$ | 760 cm$^{-1}$/94.2 meV |
| $\omega_{LO,R}$ | 1614 cm$^{-1}$/200.1 meV | $\omega_{LO,z}$ | 825 cm$^{-1}$/102.2 meV |
| $\gamma_R$ | 7 cm$^{-1}$/0.87 meV | $\gamma_z$ | 3 cm$^{-1}$/0.37 meV |

**Supplementary Table 1.** Values of the parameters entering the model dielectric tensor describing the naturally abundant hBN samples studied in experiments.

Analytical Modeling of EELS: Excitation of Guided Bulk Hyperbolic Polaritons by an Electron Beam

A very relevant model for comparison with our experimental results is that of an electron beam passing through an infinitely extended thin film of hBN following a trajectory described by the vector $(0,0,z)$. In the following, we outline a derivation of the induced electromagnetic field inside and outside the film, analyze the nature of the corresponding excitations, and evaluate the electron energy-loss probability in a way analogous to previous work[7].

By taking advantage of the rotational symmetry of the problem, we can write the wave equation for the electric field components, in particular the radial component $E_R(q,z)$ ($E_R^2 = E_x^2 + E_y^2$) and the out-of-plane component $E_z(q,z)$. We use the Fourier transform from real-space coordinates to the reciprocal space ($R \to q$, where $R^2 = x^2 + y^2$) to write the wave equation for the two non-zero components of the electric field $(E_R, E_z)$ as

$$\left(-q^2 + \frac{\partial^2}{\partial z^2}\right)(E_R, E_z) - \left(iq, \frac{\partial}{\partial z}\right)\left(iqE_R + \frac{\partial E_z}{\partial z}\right) = -\mu_0 i\omega(0, j_z) - \frac{\omega^2}{c^2}(\epsilon_R E_R, \epsilon_z E_z), \quad (8)$$

where $\mu_0$ is the vacuum permeability and $j_z = -e\, e^{i\omega z/v}$ is the free current density produced by the fast electron of elementary charge -e and speed $v$. In the above equation, we set $\epsilon_R$ and $\epsilon_z$ unequal and different from 1 when solving for the electric field inside the slab placed between $z = -t/2$ and $z = t/2$, and $\epsilon_R = \epsilon_z = 1$ elsewhere.

If we combine the wave equation, Supplementary Eq. (8), with the Fourier-transformed Gauss law

$$\frac{\partial E_z}{\partial z} = \frac{\rho_F}{\epsilon_0 \epsilon_z} - \frac{\epsilon_R}{\epsilon_z}(iq E_R), \quad (9)$$

where $\rho_F = -e/v\, e^{i\omega z/v}$ is the free charge density and $\epsilon_0$ is the vacuum permittivity, we can easily retrieve a solution for the electric field component $E_R$ inside the slab:

$$E_R\left(q, |z| < \frac{t}{2}\right) = A e^{\alpha z} + B e^{-\alpha z} + \frac{iqe\, e^{i\omega z/v}}{v\epsilon_0 \epsilon_z \left(\frac{\omega^2}{v^2} + \alpha^2\right)}, \quad (10)$$

where $A$ and $B$ are integration constants and we define $\alpha^2 = \epsilon_R \left(\frac{q^2}{\epsilon_z} - \frac{\omega^2}{c^2}\right)$. The $z$ component of the electric field can be obtained by plugging Supplementary Eq. (10) in Eq. (9) and performing the integration over $z$:

$$E_z\left(q, |z| < \frac{t}{2}\right) = -i\frac{q\epsilon_R}{\alpha \epsilon_z}[A e^{\alpha z} - B e^{-\alpha z}] + \left(1 - \frac{\epsilon_R}{\epsilon_z}\frac{q^2}{\frac{\omega^2}{v^2}+\alpha^2}\right)\frac{ie\, e^{i\omega z/v}}{\epsilon_0 \epsilon_z \omega}. \quad (11)$$

The electric field outside the slab is easily obtained from Supplementary Eqs. (10) and (11) by setting $\epsilon_R = \epsilon_z = 1$ and further requiring that the field components vanish at infinity:

$$E_{>,R}\left(q, z > \frac{t}{2}\right) = B_> e^{-\alpha_0 z} + \frac{iqe\, e^{i\omega z/v}}{v\epsilon_0 \left(\frac{\omega^2}{v^2}+\alpha_0^2\right)},$$

$$E_{<,R}\left(q, z < -\frac{t}{2}\right) = A_< e^{\alpha_0 z} + \frac{iqe\, e^{i\omega z/v}}{v\epsilon_0 \left(\frac{\omega^2}{v^2}+\alpha_0^2\right)},$$

$$E_{>,z}\left(q, z > \frac{t}{2}\right) = i\frac{Q}{\alpha_0} B_> e^{-\alpha_0 z} + \left(1 - \frac{q^2}{\frac{\omega^2}{v^2}+\alpha_0^2}\right)\frac{ie\, e^{i\omega z/v}}{\epsilon_0 \omega},$$

$$E_{<,z}\left(q, z < -\frac{t}{2}\right) = -i\frac{Q}{\alpha_0} A_< e^{\alpha_0 z} + \left(1 - \frac{q^2}{\frac{\omega^2}{v^2}+\alpha_0^2}\right)\frac{ie\, e^{i\omega z/v}}{\epsilon_0 \omega},$$

where we define $\alpha_0^2 = q^2 - \frac{\omega^2}{c^2}$ and introduce constants $A_<$ and $B_>$. The constants $\{A, B, A_<, B_>\}$ are determined by applying the boundary conditions for the field components at the slab interfaces. In particular, we require

$$E_{>,R}\left(z = \frac{t}{2}\right) = E_R\left(z = \frac{t}{2}\right), \quad E_{<,R}\left(z = -\frac{t}{2}\right) = E_R\left(z = -\frac{t}{2}\right),$$
$$E_{>,z}\left(z = \frac{t}{2}\right) = \epsilon_z E_z\left(z = \frac{t}{2}\right), \quad E_{<,z}\left(z = -\frac{t}{2}\right) = \epsilon_z E_z\left(z = -\frac{t}{2}\right). \quad (12)$$

The solutions for the field components $E_z(q, z)$, $E_{>,z}(q, z)$, and $E_{<,z}(q, z)$ are then used to evaluate the loss probability $\Gamma(\omega)$ corresponding to the experimentally measured spectra. We find that the spectral dependence can be written as

$$\Gamma(\omega) = \int_0^{q_c} dq\; P(q, \omega), \quad (13)$$

where $q_c$ is the cutoff value for the perpendicular wave vector $q$ (approximately related to the maximum collection angle of the spectrometer), and we define the momentum dependent loss probability $P(q, \omega)$ as

$$P(q, \omega) = \frac{eq}{2\pi^2 \hbar \omega}\, \mathrm{Re}\left[\int_{-\infty}^{\infty} dz\; E_z(q, z) e^{-i\omega z/v}\right],$$

which, after substitution of the electric field, yields

$$\Gamma(\omega) = \Gamma_{\text{bulk}}(\omega) + \Gamma_{\text{guid+Begr}}(\omega).$$

Here, the bulk contribution (as if the electron were moving in infinite bulk hBN) is given by

$$\Gamma_{\text{bulk}}(\omega) = \int_0^{Q_c} dq \; P_{\text{bulk}}(q,\omega) = -\frac{te^2}{2\pi^2\hbar\omega^2\epsilon_0}\text{Im}\left[\int_0^{q_c} dq \; \frac{q}{\epsilon_z}\left(1 - \frac{\epsilon_R}{\epsilon_z}\frac{qQ^2}{\frac{\omega^2}{v^2}+\alpha^2}\right)\right] = -\frac{te^2}{4\pi^2\hbar v^2\epsilon_0}\text{Im}\left[\left(\frac{1}{\epsilon_R} - \beta^2\right)\ln\left(\frac{q_c^2 v^2 \epsilon_R}{\omega^2\epsilon_z(1-\beta^2\epsilon_R)}+1\right)\right], (14)$$

where $\beta = v/c$, and the contribution emerging in the presence of the slab boundaries is given by

$$\Gamma_{\text{guid+Begr}}(\omega) = \int_0^{q_c} dq \; P_{\text{guid+Begr}}(q,\omega) = \frac{v}{2\pi^2\hbar\omega}\text{Im}\left[\int_0^{q_c} dq \; q^2 \left\{\frac{2\epsilon_R}{\epsilon_z\alpha}\left(\frac{A\sinh\left(\frac{t}{2}(\alpha-i\omega/v)\right)}{\alpha v - i\omega} - \frac{B\sinh\left(\frac{t}{2}(\alpha+i\omega/v)\right)}{\alpha v + i\omega}\right) + \frac{1}{\alpha_0}\left(\frac{A_<\exp\left(\frac{t}{2}(-\alpha_0+\frac{i\omega}{v})\right)}{\alpha_0 v - i\omega} - \frac{B_>\exp\left(-\frac{t}{2}(\alpha_0+\frac{i\omega}{v})\right)}{\alpha_0 v + i\omega}\right)\right\}\right]. (15)$$

More explicit expressions can be obtained after evaluating the coefficients using Supplementary Eqs. (12).

We concentrate now on the maxima of the probabilities, which arise as the poles of the loss functions $P(q,\omega)$. We first note that the poles of $P_{\text{bulk}}(q,\omega)$ are identical to Supplementary Eq. (7) with $q_z = \omega/v$ matching the wave vector provided by the fast electron in the $z$ direction, and thus, we confirm that the electron can launch waves in bulk hBN. By analyzing poles of $P_{\text{guid+Begr}}$, we find a contribution that we denote as the Begrenzungs term, which is responsible for the compensation of the bulk losses and therefore has to have the same denominator as the $P_{\text{bulk}}$ in Eq. (9). As the slab gets thinner, the Begrenzungs contribution tends to cancel the bulk contribution $\Gamma_{\text{bulk}}$. In addition, the poles corresponding to the guided loss $P_{\text{guid}}$ are given by the following transcendental equations

$$\frac{\alpha}{\epsilon_R\alpha_0}\coth(\alpha t) + 1 = 0, \quad (16a) \qquad \frac{\alpha}{\epsilon_R\alpha_0}\tanh(\alpha t) + 1 = 0. \quad (16b)$$

We find that Supplementary Eq. (16a) gives the dispersion of p-polarized charge-symmetric slab phonon-polariton modes (with respect to $z = 0$) and Supplementary Eq. (16b) corresponds to charge-antisymmetric modes, which propagate along the slab surface[8]. Due to the hyperbolic nature of isofrequency surfaces of hBN, these modes are denoted as bulk HPhPs. In the quasistatic limit ($c \to \infty$), which works very well in the upper RB and for relatively thin hBN slabs ($d \lesssim 100$ nm), the dispersion of bulk HPhPs can be expressed by a single equation[9]

$$Qt = \text{Re}\left[i\sqrt{\frac{\epsilon_z}{\epsilon_R}}\left\{\pi n + 2\,\text{atan}\left(\frac{i}{\sqrt{\epsilon_R\epsilon_z}}\right)\right\}\right], (17)$$

where $n = 0,1,2,...$ is the mode order. For hBN, we have symmetric and antisymmetric modes for $n$ even and odd, respectively. We highlight that hBN can sustain in its RBs infinite number of symmetric and antisymmetric slab modes analogous to those of a dielectric slab waveguide. However, in practice, the contribution of the lowest-order mode ($n = 0$) is dominant over the higher-order modes. At $n=0$, Supplementary Eq. (17) reduces to the analytical expression used to represent the bulk polariton in the main text, Supplementary Eq. (1). In Main Text Fig. 2e, we plot the dispersion of this mode (solid red curve) on top of the experimental dispersion.

In contrast, if we had a slab of an isotropic material, we would observe only one symmetric and one antisymmetric mode inside the RB. These two modes (Fuchs-Kliewer modes) emerge from the coupling of two surface waves propagating on the top and bottom boundary of the hBN slab.

Our analysis shows that the fast electrons naturally launch bulk polariton modes propagating in hBN slabs. The excitation of these modes is also reflected in the integrated EELS probability $\Gamma(\omega)$. The integration over the perpendicular momenta according to Supplementary Eq. (13) can be performed analytically only for the bulk term. The guided and Begrenzungs loss terms have to be integrated numerically, in practice up to a finite cutoff value, which has to be much smaller than the inverse of the distances between atoms, and therefore, it is reasonable to integrate only up to $Q_c \approx 1$ Å$^{-1}$. An example of the integrated spectrum is shown in Main Text Fig. 4h (black curve). We find that it exhibits two peaks: (i) a broad peak around 172 meV, which originates dominantly from the fundamental charge-symmetric polariton mode [with dispersion given by Supplementary Eq. (16a), taking the solution with the smallest $Q$ for a fixed $\omega$, or by setting $n = 0$ in Supplementary Eq. (17)]; and (ii) a narrow peak close to $\omega_{LO,R}$, which arises from the bulk loss term $\Gamma_{bulk}$ [Supplementary Eq. (14)]. The latter peak would be present also if the electron were moving in unbounded hBN.

Analytical Modeling of EELS: Excitation of Edge-Guided Hyperbolic Polaritons by an Electron Beam
To model the edge polariton, we restrict ourselves to non-retarded analysis, which works very well for this geometry and in the upper RB or hBN. An extended analysis for an electron beam interacting passing in aloof geometry with respect to an infinite interface has been previously reported[5]. If we assume an infinite interface in the plane $x = 0$ between hBN (in $x < 0$ half-space) and vacuum (in $x > 0$ half-space), by utilizing the translational symmetry in the $y$ direction and modifying the Poisson equation ($\nabla \cdot (\hat{\epsilon} \nabla \Phi) = \rho_F/\epsilon_0$), after some algebra, we arrive to solution for the electrostatic potential $\Phi$ in hBN and vacuum:

$$\Phi_>(x > 0, q_y, q_z, \omega) = Ae^{-\lambda_0 x} - \frac{\pi e \delta\left(q_z - \frac{\omega}{v}\right)}{\epsilon_0 v \lambda_0} e^{-\lambda_0 |x - x_b|}, \qquad \Phi_<(x < 0, q_y, q_z, \omega) = Be^{\lambda x},$$

where we impose the vanishing of the potential at infinity, the beam is passing at $y_b = 0$ and distance $x_b > 0$ from the interface along the $z$ axis, and we have defined $\lambda_0^2 = q_z^2 + q_y^2$, $\lambda^2 = \frac{q_z^2 \epsilon_z}{\epsilon_R} + q_y^2$. We take hBN to be aligned in such a way that the $\epsilon_z$ is parallel with the $z$ direction (and the beam axis). The coefficients $A$ and $B$ are determined from the boundary conditions at $x = 0$
$\Phi_>(x = 0) = \Phi_<(x = 0)$ and $\left.\frac{d\Phi_>}{dx}\right|_{x=0} = \epsilon_R \left.\frac{d\Phi_<}{dx}\right|_{x=0}$.
Finally, we obtain

$$A = \frac{\pi e \delta\left(q_z - \frac{\omega}{v}\right)}{\epsilon_0 v} e^{-\lambda_0 x_b} \frac{\epsilon_R \lambda - \lambda_0}{\lambda_0(\epsilon_R \lambda + \lambda_0)}, \qquad B = \frac{-2\pi e \delta\left(q_z - \frac{\omega}{v}\right)}{\epsilon_0 v (\epsilon_R \lambda + \lambda_0)} e^{-\lambda_0 x_b},$$

and the loss probability per unit trajectory can in turn be expressed as

$$\frac{d\Gamma(\omega)}{dz} = \frac{2e}{(2\pi)^3 \hbar \omega} \int_{-\infty}^{\infty} dq_z \int_{-\infty}^{\infty} dq_y \, \text{Im}\left[q_z \Phi_>(x_b, q_y, q_z, \omega) e^{i z(q_z - \omega/v)}\right]$$

$$= \frac{e^2}{2\pi^2 \hbar v^2 \epsilon_0} \int_0^{\infty} dq_y \, \frac{e^{-2\lambda_0 x_b}}{\lambda_0'} \text{Im}\left[\frac{\epsilon_R \lambda' - \lambda_0'}{\epsilon_R \lambda' + \lambda_0'}\right],$$

where $\lambda_0' = \sqrt{\frac{\omega^2}{v^2} + q_y^2}$ and $\lambda' = \sqrt{\frac{\epsilon_z \omega^2}{\epsilon_R v^2} + q_y^2}$. The poles of the loss probability given by $\epsilon_R \lambda' + \lambda_0' = 0$ again lead to a dispersive resonant behavior, which, upon inspection the field profile, we assign to an edge HPhP that is strongly localized at the interface (at $x = 0$) and propagates in this plane[6]. We find that for $q_y \to \infty$, the dispersion of the mode reaches limiting energy when $\epsilon_R(\omega_{Edge-PhP}) + 1 = 0$, which for naturally abundant hBN happens at $\hbar\omega_{Edge-PhP} = 195$ meV.

To find the dispersion relation of the polaritons guided along the edge (truncation) surface, we apply a Fabry-Perot model assuming reflection of the mode at the boundaries of the surface at $z = \pm t/2$. By further comparison with numerical results, as the phase describing reflection from the surface boundaries is unknown, we obtain an approximate analytical expression for the dispersion of the fundamental edge-guided polariton mode, Supplementary Eq. (4), $q_y = \pi/t \, \text{Re}\left[\frac{1}{\epsilon_z} \frac{\epsilon_R \epsilon_z - 1}{1 - \epsilon_R^2}\right]$.

Modeling the EELS Probability Close to hBN Corner
The phenomenological reflection model can be applied to model the reflection of bulk polaritons off two edges as involved in the "corner" geometry (see the schematics in Main Text Fig. 5) yielding

$$\Gamma_{corner,2D}(\omega, x_b, y_b) \approx \int_0^{Q_c} dQ \; P(Q, \omega) \left[2 + \cos(2Q[\cos\beta(\tan\beta \; y_b + x_b)] + \phi_R) + \cos(2Q[\cos\beta(\tan\beta \; y_b - x_b)] + \phi_R)\right], \quad (18)$$

where $\beta$ is half-angle at the corner tip and $P = P_\text{bulk} + P_\text{guid+Begr}$ is defined in Supplementary Eqs. (14) and (S15). Notice that the contribution of the two reflected HPhP waves is added as a sum of two independent contributions, because the wave vectors of the two reflected waves are not parallel at the beam position and thus do not interfere. In Supplementary Figure S19, we plot the resulting energy-filtered EELS map at 190 meV, which can be compared with fully numerical calculation in Main Text Fig. 5. As expected, the pattern further from the edges (where we cannot disregard the contribution of edge polaritons) is relatively well reproduced even with this simple model.

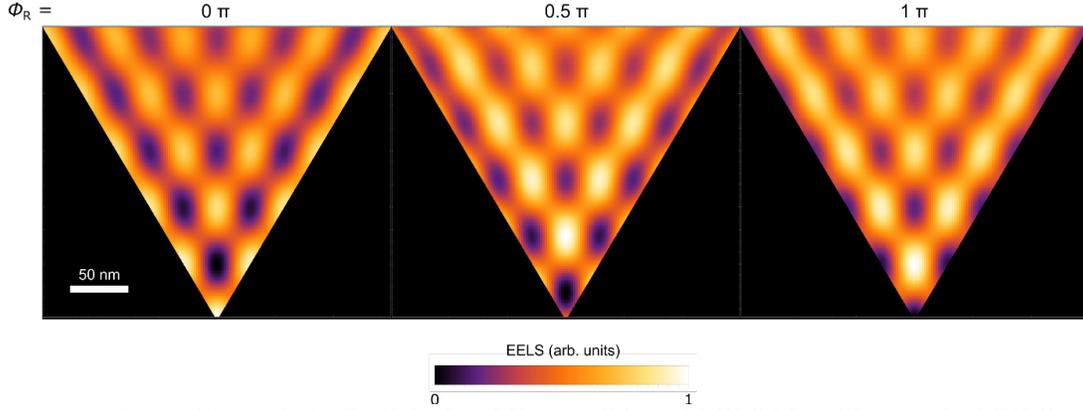

**Supplementary Figure 19. Analytically Calculated Energy-Filtered EELS Map Close to the hBN Corner.** We use Supplementary Eq. (18), considering an energy of 190 meV, $\beta = 30°$, and a slab with thickness of 12 nm. We vary $\phi_R$ as denoted above each panel.

Localized Phonon-polariton Modes in hBN Disks
Modes of a finite hBN nanodisk can be understood in a similar manner as those of the hBN triangle: there are modes formed along the nanostructure edges and across its volume. The former class of modes is responsible for lower-energy peaks in the EELS spectra shown in Main Text Fig. 4g. In particular, the modes close to 171.7 meV and 173.8 meV can be classified as dipolar and quadrupolar, respectively, and exhibit the corresponding charge variation along the disk circumference, as it was shown in Supplementary Fig. 4b and 4c earlier in the Supplementary Information.

In addition, the fundamental "volume" mode is a breathing mode appearing close to 177 meV (Supplementary Fig. 4c). This mode has a net dipole moment and is efficiently excited for the beam placed at the nanodisk center. Higher-order "volume" modes exhibit additional charge variation with the azimuthal angle, with an example of the dipole breathing mode close to 179.5 meV (Supplementary Fig. 4d). These modes are analogous to those that are observed in isotropic disks supporting plasmon polaritons[10], or phonon polaritons hBN nanoantennas[1].

Additional Detail on Numerical Calculations
We used an implementation of the finite element method in the commercial software Comsol Multiphysics. We utilized the Radio Frequency Toolbox to perform retarded simulations (Maxwell's equations are solved) or the AC/DC Toolbox for non-retarded simulations (Poisson's equation is solved) to evaluate the electric field in presence of hBN nanostructures and a current source representing an electron beam aligned along the $z$ direction, expressed as $I = I_0 e^{i\omega z/v}$, where $I_0$ is the amplitude of the electric current, $\omega$ is the angular frequency, and $v$ is the electron speed. In all calculations, we used $v = 0.328\,c$, corresponding to 30-keV electrons.

After the solution for the electric field was obtained, we evaluated the loss probability $\Gamma$ (using *Edge Probe, Integral* applied along the electron trajectory) according to

$$\Gamma(\omega) = e^2/(\pi\hbar\omega I_0) \int_{z_\text{min}}^{z_\text{max}} dz\, \text{Re}\{E_z(x_\text{b}, y_\text{b}, z) e^{-i\omega z/v}\}, \quad (19)$$

where $e$ is the elementary charge, $\hbar$ is the reduced Planck constant, $E_z$ is the $z$ component of the electric field, and we considered the electron beam trajectory intersecting the $xy$ plane at the position $(x_\text{b}, y_\text{b})$. The simulation domain was truncated at coordinates $z_\text{min/max}$ in the $z$ direction.

First, we performed simulations with the dielectric response modeled by the tensor $\hat{\epsilon}_{\text{hBN}}(\omega)$ (see Supplementary Table 1 for values we used to model the anisotropic dielectric response of hBN) filling regions of the hBN nanostructures and $\epsilon = 1$ elsewhere, from which we obtain the loss probability $\Gamma_1$. Next, we set $\epsilon = 1$ everywhere, so that only the field of the electron is present, preserving the same discretization (meshing) of the geometrical domains, and calculated the loss probability $\Gamma_2$. To obtain only the contribution of the induced field coming from the interaction of the electron beam with the nanostructure and to correct for the finite length of the electron trajectory $|z_{\max} - z_{\min}|$ and non-zero values of the fast electron field close to the boundaries of the simulation domain, the loss probability due to the induced near field that we seek for was obtained as $\Gamma_{\text{corr}} = \Gamma_1 - \Gamma_2$.

For the hBN edge, we took advantage of translational symmetry along the $y$ direction (see schematics in Main Text Figure 3) and performed the calculations in 2D simulation domain, where the solution of the electric field can be written as $E_z(x,z)e^{iq_y y}$, where $q_y$ is the out-of-plane wave vector (in the direction of the translational symmetry). Similarly, the applied line current source can be decomposed as $I = I_0 e^{i\omega z/v} e^{iq_y y}$. Supplementary equation (19) is then modified to

$$\Gamma(\omega) = 2e^2/(\pi\hbar\omega I_0) \int_{z_{\min}}^{z_{\max}} dz \int_0^{q_{y,c}} dq_y \, P(x_b, q_y, \omega) \, , (20)$$

where we evaluated the field at $y_b = 0$ and where $q_{y,c}$ is an out-of-plane wave vector cutoff. We also defined the probability $P(x_b, q_y, \omega) = \text{Re}\{E_z(x_b, q_y, z)e^{-i\omega z/v}\}$. In practice, we selected the cutoff and a discrete step in $q_y$ in such a way that the numerically summed spectrum was converged. Similarly, by summing over $q_y$, we could retrieve the electric field plots.

We used standard Comsol meshing options in 3D and 2D. In particular, *Free Tetrahedral*/*Free Triangular* elements. We applied meshing with refined elements in areas of high field concentration and gradients, typically close to the electron trajectory and close and inside the hBN nanostructures. We allowed for an increase of the size of the mesh elements towards outer boundaries of the simulation domain. Each simulation domain was truncated by *Perfectly Matched Layers* (PML) domain that helped to attenuate the electric field at the boundaries of the simulation domain and prevented unphysical field reflection from the boundaries. The PML domain was meshed by 5-10 *Swept* layers. The maximal allowed element dimensions depended on the simulated energy region and thus on the typical wavelengths involved. We typically used fractions of the typical wavelength for the largest elements.